\documentclass[reprint, superscriptaddress, amsmath,amssymb, aps, prx, longbibliography]{revtex4-2}

\usepackage{graphicx}
\usepackage{multirow}
\usepackage{comment}
\usepackage[usenames]{color}
\usepackage{graphicx}
\usepackage{dcolumn}
\usepackage{bm}
\usepackage[colorlinks=true, allcolors=blue]{hyperref}
\usepackage{amssymb,amsfonts,amsmath}

\begin{document}

\author{Ezequiel E. Ferrero} 
\affiliation{Instituto de Nanociencia y Nanotecnolog\'{\i}a, CNEA--CONICET, 
Centro At\'omico Bariloche, (R8402AGP) San Carlos de Bariloche, R\'{\i}o Negro, Argentina.}
\affiliation{Departament de Física de la Matèria Condensada, Universitat de Barcelona, Martí i Franquès 1, 08028 Barcelona, Spain.}
\affiliation{Institute of Complex Systems (UBICS), Universitat de Barcelona, Barcelona, Spain
}

\author{Eduardo A. Jagla} 
\affiliation{Centro At\'omico Bariloche, Instituto Balseiro, 
Comisi\'on Nacional de Energ\'ia At\'omica, CNEA, CONICET, UNCUYO,\\
Av.~E.~Bustillo 9500 (R8402AGP) San Carlos de Bariloche, R\'io Negro, Argentina.}

\title{
Soil creep facilitated by cyclic variations of environmental conditions
}

\begin{abstract} 
Sloped terrains tend to creep downward over time, even when their slope 
is below the nominal angle of repose. 
This behavior can result from periodic variations in environmental conditions, 
such as daily or seasonal fluctuations in temperature and humidity.
We study this process by considering a model of an athermal yield stress material 
under an applied stress lower than the critical yield stress value $\sigma_c$. 
Normally, in such a
situation the material does not flow at all. 
However, under cyclic temporal variation of system parameters
a finite amount of irreversible deformation
can remain after each cycle, 
and a long term steady-state flow of the whole system can be induced.
In our model, we cycle the strength of internal elastic interactions 
to mimic the effect of cyclic variation of environmental conditions in the real soils.
We find that the amount of deformation per cycle increases if 
$\sigma_c$ is approached from below, and it decreases and even vanishes at 
a novel critical stress $\sigma_0<\sigma_c$ when this, in turn, is reached from above.
Interestingly, $\sigma_0$ plays a role similar to the endurance limit in the context 
of fatigue damage propagation. 
Despite the model's simplicity, our results offer a fresh perspective on subcritical 
landform evolution, with implications for the creep of hill slopes over long periods 
and the precursors to runaway landslides.

\end{abstract}

\maketitle

\section{Introduction}
\vspace{-0.2cm}

There is a renewed interest in the study of the `thin skin' of the Earth~\cite{daniels2019viewing,VoiglanderSM2024}.
The understanding of the soft matter landscape on which we live comes up
as increasingly essential in times of climate change.
Several hazardous events, such as landslides, earthquakes, faulting, and ice fractures, 
are related to the slow evolution of landscapes.
In particular, well-known evidence indicates that terrains systematically evolve 
downhill over long periods of time (years and beyond), a phenomenon known as \textit{soil creep}
in geophysics~\cite{YoungNature1960,KirkbyIBGSP1971}.
This sub-critical crawling motion exhibits dynamics similar to that of yield-stress materials.
These are systems encompassing gels, foams, emulsions and polymeric, colloidal and 
granular glasses, characterized by a macroscopic persistent deformation rate if 
applied stress $\sigma$ is larger than some critical 
value $\sigma_c$~\cite{BonnRMP2017,NicolasRMP2018}.
This analogy has motivated the study of geophysical problems with tools and models
inherited from condensed matter and statistical physics, on a field now called
Soft Earth Geophysics~\cite{daniels2019viewing,VoiglanderSM2024}.

Under sub-critical conditions ($\sigma<\sigma_c$) the deformation of amorphous
materials can be either a transient effect (usually referred to as `Andrade 
creep')~\cite{BonnRMP2017,NicolasRMP2018,PopovicPRL2022,LiuSM2018,WeissPRM2023}
or a thermally activated flow~\cite{PopovicPRE2021, FerreroPRM2021},
which eventually at very small driving is analogous to the 
thermal creep of elastic interfaces in random 
media~\cite{ChauvePRB2000, FerreroARCMP2021, ferrero2017spatiotemporal}.
In soils (a case of granular matter composed by sand, rocks, clay, organic remains, etc.), 
it is quite clear that thermally activated processes are almost negligible, 
and the possibility of long lasting but transient 
deformations is under debate. 
For instance, recent experiments on sand-piles~\cite{DeshpandeNC2021,Deshpande2024arXiv},
essentially an \textit{athermal} system, have shown sustained creep motion at 
sub-critical slopes in undisturbed setups, and presented these 
``quenched quiescent heaps that creep indefinitely'' as a challenge to 
granular rheology.
This raises a natural question: \textit{What other sub-critical flow mechanisms, aside 
from transient or thermally activated creep, should be considered in Soft Earth Geophysics?}

Unlike typical soft matter systems studied under controlled laboratory conditions, 
soils experience various mechanical perturbations that, along with gravity, can 
contribute to sub-critical flow~\cite{FerdowsiPNAS2018, HoussaisNatComm2015, JaglaSM2023}.
They include vibrations caused by walking of animals, 
vegetation movement due to wind, water falling and flow during rain, and
even earthquakes. 
To some extent, all of them have the potential to produce a persistent down hill 
evolution of the soil~\cite{EylesJTG1970,FlemingQJEGH1975,MatsuokaPPP1998,AuzetESPL1996,BontempsNC2020}.
Other relevant source of external perturbations comprise those 
originated by periodic variations of parameters through changes 
of environmental conditions that affect the internal properties of the system. 
Key examples include daily or seasonal variations in temperature and 
humidity, which microscopically alter the size, surface properties, 
and mechanical response of soil constituents, thereby affecting the 
internal interactions and evolution of the system.
As a matter of fact, thermal cycling effects (sometimes referred to 
as thermo-mechanical ratcheting) have been studied in granular 
systems in the last years~\cite{DivouxPRL2008, PercierEPL2013, 
PastenAG2019, CoulibalyGM2020, RottaLoriaGEE2021, PanGM2024, PanSR2024},  
primarily by the mechanical engineering community.
Overall, there is qualitative agreement that oscillatory changes in 
environmental conditions can significantly affect the dynamic evolution 
of the system.
In the case of a sloped terrain, this may lead to a persistent 
downhill displacement.
This phenomenon is the focus of the present manuscript, where we 
mainly engage with models and ideas from driven phase transitions 
in disordered systems.

The first report of the phenomenon of downhill movement under 
oscillatory external conditions is most likely that of Moseley~\cite{MoseleyPRSL1856} 
in his letter `On the descent of glaciers' (1856).
He analyzed the case of a slab of material resting on an inclined plane 
due to frictional contact. 
Under periodic variations of temperature, the slab expands and contracts, 
and a simple mechanical analysis predicts that on each full variation cycle
there is a net descent of the slab. 
His idea was criticized at that time for being over-simplistic.
Nevertheless, it describes the essential phenomenology that is present 
in more complex and realistic systems.
Moseley's idea was recapitulated more recently 
by  Croll~\cite{CrollPRSL2009}, who discussed with illustrative examples 
of ice-rich materials and asphalt pavements that, when a solid is subject
to alternations of tension and compression (following alternations in temperature), 
some motion can be produced even in situations where gravity is either absent or 
further against the prospective motions.
Blanc, Pugnaloni and G\'eminard~\cite{BlancPRE2011} have applied the analysis 
of Moseley to a one dimensional chain of blocks connected through 
elastic springs that rest on an incline.
Introducing a cyclic variation of the rest length of the springs 
(mimicking a thermal expansion-contraction of a macroscopic material) 
they observed a reptation of the chain down-hill and were able to estimate 
its average creep velocity. 
Notably, they used a phenomenological Amontov-Coulomb friction law between blocks 
and the substrate. Additionally, the absence of stochastic elements in the model 
led to behavior reminiscent of ideal dynamical systems, such as peculiar 
synchronizations, limit cycles, and plateaus in the dynamical evolution.
Although with limitations, these previous works already gave a qualitative idea
of the phenomenon we will discuss: a sub-critical flow based on the periodic 
variation of inter-element interactions that we can ascribe in real systems 
to changes of environmental conditions.

Our approach introduces some elements that bring these ideas closer to the 
effective description of the concrete phenomenon of soil creep.
First of all, we do not introduce any \textit{ad hoc} form for a friction law. 
Instead, we consider the overdamped evolution of a system of 
mesoscopic `blocks' or regions of an amorphous material, 
and eventually the appearance of a friction-like law 
(viz., depinning/yielding) is an emergent property in our treatment.
Our model incorporates a degree of randomness, coded mainly by 
the stochastic form of the interaction/deformation potentials, 
and this smooths out the synchronization effects that might 
appear in the absence of such a randomness.
Finally and most importantly, we do not limit to the description 
of a frictional situation between two solid bodies,
neither to zero- or one-dimensional systems.
In fact, we consider here two-dimensional systems of two 
families of problems: 
(i) {\em depinning models}, typically used to describe the driven transition 
between rest and movement of an elastic manifold driven on top of a disordered 
pinning potential, and 
(ii) {\em yielding models}, typically used to describe the bulk deformation of 
an homogeneous amorphous material under an applied external shear stress.

In particular, the implementation of  the yielding case is suited to describe 
the slow deformation of a bulk material, which is the case that is most relevant 
to describe the soil creep phenomenon.
Returning to our inspiration from Soft Earth Geophysics 
problems~\cite{daniels2019viewing, DeshpandeNC2021, Deshpande2024arXiv}, 
we believe that the kind of sub-critical deformation mechanism that we 
analyze, its underlying mechanisms and universal characteristics, can be further 
explored and extended to tackle concrete examples in that field.

To provide insights in the basic underlying phenomenology, 
in Sec.~\ref{sec:twoparticles} we first present the simple example of a 
two-particle system joined by a spring that changes its stiffness in a 
periodic way.
Then, we introduce our modeling framework (Sec.\ref{sec:framework})
and study the oscillatory creep phenomenology in two spatially 
extended two-dimensional models:
(i) an elastoplastic model of amorphous solids, with long-range elastic 
interactions in Sec.~\ref{sec:yieldingcreep},
and (ii) a driven elastic interface with short range elastic interactions,
in Sec.~\ref{sec:depinningcreep}. 
In Sec.~\ref{sec:meanfieldcreep}, we formalize these findings considering 
a fully-interacting mean-field system where some results can be deduced 
analytically.
Finally, we present our conclusions and leave some open questions 
in Sec.~\ref{sec:discussion}. 
In Appendix~\ref{app:analyticMF} we derive the analytic results of the 
mean-field model presented in Sec.~\ref{sec:meanfieldcreep}.

\section{Reptation of a two-particle system caused by oscillation of the interaction intensity}
\label{sec:twoparticles}

\begin{figure}[t!]
\includegraphics[width=0.9\columnwidth,clip=true]{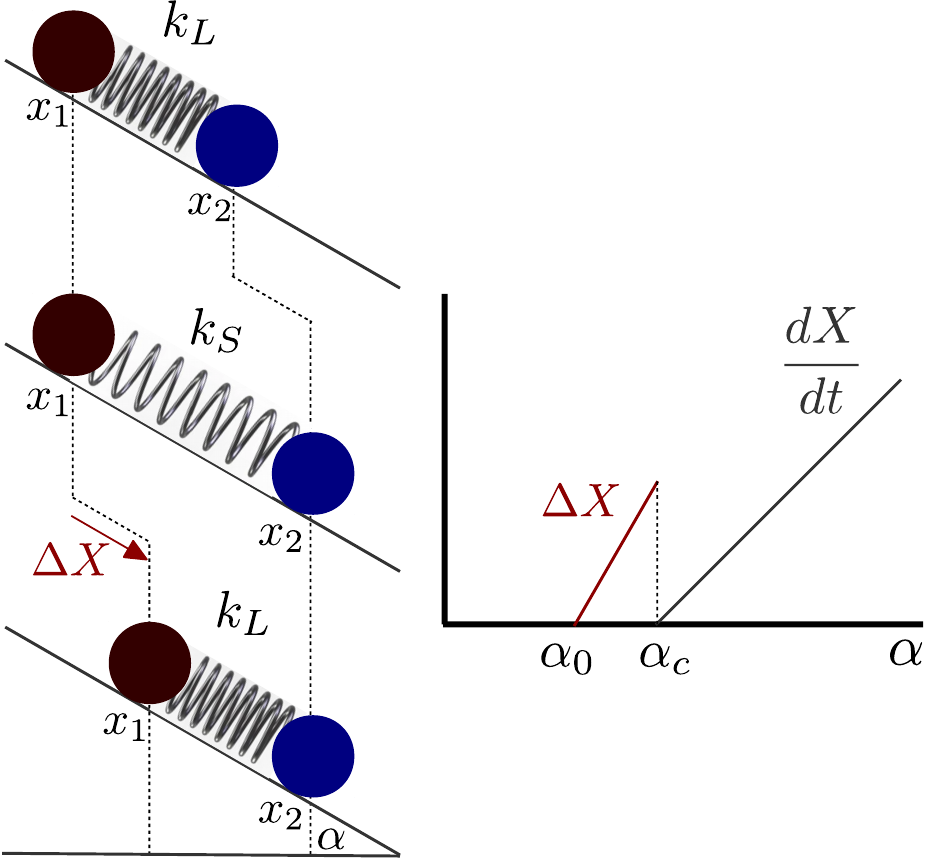}
\caption{
\textit{Schematics of the reptation process by cyclic variations
in the interaction forces: }
Two particles linked by a spring of strength $k$ rest on an inclined.
For constant $k$ the system behaves according to the black line in 
the right plot that shows displacement rate $dX/dt$ vs. 
incline angle $\alpha$, $\alpha_c$ being the minimum slope for system movement.
If $\alpha<\alpha_c$, the system performs a reptation
upon changing the value of $k$ between $k_L$ and $k_S$,
advancing a distance $\Delta X$ on each cycle.
There is a minimum slope $\alpha_0$ for this process to occur.
}
\label{fig:two-particle-system}
\end{figure}

We  analyze an elementary system that qualitatively displays the 
essence of the physical process under study~\cite{BlancPRE2011}.
Let us consider two particles of mass $m$ joined by a spring,
lying on a slope. 
For a fixed value of the spring constant $k$, the system may be at rest, or 
smoothly sliding depending on the value of the slope angle, and the critical 
friction forces of the particles. 
Assuming there is some asymmetry between the particles such that the critical 
friction forces of each of them are $f_1$ and $f_2$, the critical
angle $\alpha_c$ for smooth descent at constant velocity is obtained as

\begin{equation}
\sin(\alpha_c)\equiv \frac{f_{1}+f_{2}}{2mg}.    
\end{equation}
Note that $\alpha_c$ is independent on the value of $k$. 

If $\alpha<\alpha_c$ we could expect that the system remains always 
at the same location.
However, if the value of $k$ fluctuates for some reason (let's say
$k$ oscillates between a large value $k=k_L$ and a small value $k=k_S$) 
and if $\alpha$ is sufficiently close to (but lower than) $\alpha_c$, 
then there is an alternate advance of $x_1$ and $x_2$
as $k$ passes from $k_L$ to $k_S$, and back to $k_L$.
This is schematically plotted in Fig.~\ref{fig:two-particle-system}. 
$\Delta X$ is the net advance of the system per cycle. 
 
The origin of this reptation phenomenon is the following. 
Starting from the configuration if Fig.~\ref{fig:two-particle-system}(a) 
with $k=k_L$, the reduction of $k$ to $k_S$ produces an increase of the force 
on the right-most particle, which moves to the right until it experiences a 
force equal to its critical force;
this is the configuration in Fig.~\ref{fig:two-particle-system}(b). 
Now, as $k$ is increased back to $k_L$, it is the left-most particle 
that receives a force larger than its critical one, and moves to the right 
until the force does not exceed the critical value any more 
(Fig.~\ref{fig:two-particle-system}(c)).
A simple calculation shows that $\Delta X=0$ below a critical angle 
$\alpha_0$ given by (we assume $f_1>f_2$)
\begin{equation}
\sin(\alpha_0)=\frac{k_Lf_{1}+k_Sf_{2}}{mg(k_L+k_S)}.
\end{equation}
If $\alpha_0<\alpha<\alpha_c$ the value of $\Delta X$ is given by
\begin{equation}
\Delta X= -\frac{f_1}{k_S}-\frac{f_2}{k_L} +mg   \left( \frac 1 {k_L}+\frac 1 {k_S}  \right)\sin(\alpha)
\end{equation}
which is indicated in blue if Fig.~\ref{fig:two-particle-system}.
This is the phenomenon we discuss in the rest of the paper, distilled to its simplest form.

\section{Theoretical framework and modeling}
\label{sec:framework}

We use a common framework for depinning and yielding phenomena, 
that of elastic manifolds evolving onto disordered energy 
landscapes~\cite{FernandezAguirrePRE2018,FerreroPRL2019,LiuJCP2022,FerreroPRM2021}.
The manifold can either represent an elastic interface
${x}({\bf r})$ that undergoes a depinning transition, 
or it can represent the local strains configuration 
${\gamma}({\bf x})$ of an amorphous material 
(in this case the energy landscape represents the 
possibility of many different locally stable configurations).
We limit ourselves to the study of two-dimensional systems of depinning 
and yielding in this work.

To fix ideas, let us first describe the case of the depinning of the elastic 
interface and then declare the analogous quantities for yielding.
Apart from the elastic interactions and the forces induced by the underlying
disorder potential, the system is subject to an external drive:
we note this forcing as $f$.
We consider the local position $x({\bf r},t)$ of an interface, that we will 
discretize on a square lattice (with periodic boundary conditions) and 
denote $x_i$ the position at site $i$. 
The temporal evolution of $x_i$ is through an overdamped dynamical equation of the form

\begin{equation}
\frac{\partial x_i}{\partial t} = - \frac{d V_i}{d x_i} + \sum_{j~\!\text{nn}\!~i} k(x_j-x_i) + f.
\label{eq:eom_depinning}
\end{equation}

\noindent where we have restricted the elastic interactions to nearest neighbors 
for simplicity. 
The disordered potentials $V_i$ are chosen to be an alternate sequence of parabolic `traps'
and flat regions, as schematically shown in Fig.~\ref{fig:disordered_potential} of 
Appendix \ref{app:methods}.
For simplicity we have chosen the traps to be identical and to encode 
the stochasticity in the length of the  flat regions of $V(x)$.
The use of random widths or depths for the parabolic wells does not change
the qualitative picture we observe, nor the critical properties of the systems
under study.

Notice that we could think both on a two-dimensional elastic manifold
advancing in the direction normal to the system coordinates or in a two-dimensional
surface sliding on top of another that plays the role of a rough substrate.
In the latter interpretation, one needs to assume that for each system's 
block the disorder explored is still independent of the one seen by the other 
blocks, but to a reasonable extent this is a good approximation and in fact this
kind of models have been used to successfully describe, for instance, 
the sliding of tectonic plates and earthquake statistics~\cite{KoltonPRE2018,Jagla2014}.

The common phenomenology to a variety of systems described by equations
similar to Eq.~\ref{eq:eom_depinning}, despite the particular kind of 
elastic interactions and disorder potentials, is the following. 
There is a critical value $f_c$, such that for $f\leq f_c$ the system 
eventually  reaches a stable 
configuration and stops evolving in time; while for $f>f_c$ it keeps evolving in a finite 
steady velocity situation.
Above and near $f_c$ the velocity $v$ of the elastic interface has a dependence on $f$ 
of the form $v\sim (f-f_c)^\beta$. 
Such a driven transitions referred to as {\it depinning}~\cite{Fisher1998,Kardar1998} 
is sharply defined only in the ideal case in which other external disturbances are 
assumed to be negligible. 
For instance, the presence of a finite temperature produces stochastic fluctuating 
forces on the elementary constituents in the system that are known to smooth out the transition, 
turning it into a crossover~\cite{BustingorryEPL2007,FerreroCRP2013,kolton2020thermally}.
This produces thermally activated creep even at very small driving 
forces~\cite{ChauvePRB2000, KoltonPRL2006, ferrero2017spatiotemporal}.

In the present work, while we stick to the athermal case, we incorporate periodic 
variations of the intensity of the elastic interactions in the system, generically 
denoted by $k$, that will produce a crawling effect on the system and a persistent evolution, 
as long as the elastic intensities continue to oscillate.
Let us recall that, for a given amplitude of the disorder potential and a fix 
strength $k$ of the elastic interactions, the critical force $f_c$ depends on $k$.
If for a value of $k=k_1$ we have a critical force $f_{c_1}$, 
for $k=k_2<k_1$ the critical force will be $f_{c_2}>f_{c_1}$. This is because
a softer elasticity allows the interface to better adapt
to the disorder potential, occupying deeper 
local energy minima, and therefore increasing the threshold force needed for depinning.
Now we consider what happens if the value of $k$ is cycled between a large value $k_L$ 
and a small value $k_S$.
Such oscillations in $k$ will produce a minor 
effect in the moving phase, $f>f_{c,k_S}>f_{c,k_L}$ since the system 
is already evolving at a finite velocity regardless the value of $k$.
The most surprising consequences of an oscillation in $k$
occur in the fully sub-critical phase $f<f_{c,k_L}<f_{c,k_S}$.
Naively one could expect no movement at all in this regime. 
However, as in problem analyzed by Moseley~\cite{MoseleyPRSL1856},
the oscillatory variation of $k$ induces a systematic advance of the elastic manifold
that is synchronized with the  cycles of the perturbation.

\begin{figure}[t!]
\includegraphics[width=0.9\columnwidth]{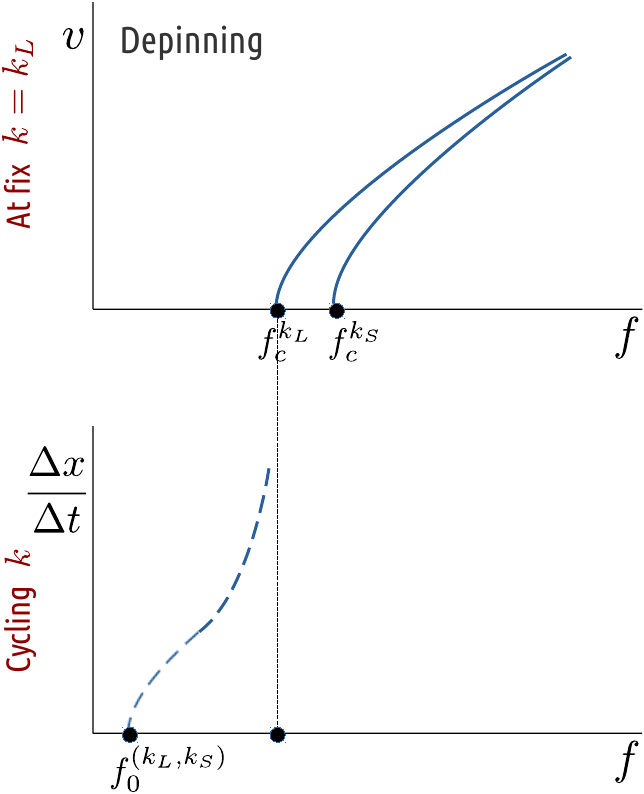}
\caption{
Schematic display of the crawling mechanism by cyclic variation of 
parameters discussed in this work.
}\label{fig:schematic_creep_depinning}
\end{figure}

In the steady state, after many oscillation cycles of $k$, an 
advance equal to $M\Delta X$ will be observed, 
where $M$ is the number of cycles applied and $\Delta X$ 
is the advance per cycle.
Even though it might be incorrect to talk about a finite \textit{velocity} 
of advance, $\Delta X$ is a clearly measurable quantity
and happens to be non-zero in a non-negligible range of sub-critical forces.
$\Delta X$ 
is particularly large when $f$ is only slightly below the critical force
$f_c=f_{c,k_L}$, and is shown to decrease as we depart from it. 
Furthermore, we are able to show that this athermal reptation assisted by 
the cyclic variation of parameters cannot happen
below a minimal external force $f_0$, with $f_0$  depending on 
$k_L$ and $k_S$.
The above dynamical scenario is schematically illustrated 
in Fig.~\ref{fig:schematic_creep_depinning}.


When switching to the case of the yielding transition of driven amorphous 
solids, we can do a complete analogy of the phenomenology described above 
for depinning.
An amorphous solid subject to an external stress $\sigma$ will
flow in the steady state if $\sigma>\sigma_c$~\footnote{In yielding under 
applied stress there is a problem similar to the one that distinguish 
static from dynamic friction. If a system is well annealed, applying a 
stress slightly above $\sigma_c$ would not be sufficient to set it in 
`motion' (plastic flow) but once it's flowing it will flow as soon as
$\sigma>\sigma_c$ is maintained. $\sigma_c$ in our framework is the dynamical
yield stress.}.
In this case, the order parameter of the transition is the deformation velocity
or strain-rate $\dot{\gamma}$ that departs from zero as 
$\dot{\gamma} \sim (\sigma-\sigma_c)^\beta$, typically with $\beta>1$.
We describe a two-dimensional material with periodic boundary conditions.
The equation of motion that we solve is now

\begin{equation}
\frac{\partial \gamma_i}{\partial t} = - \frac{d V_i}{d \gamma_i} + \sum_{j} G_{ij}\gamma_j + \sigma.
\label{eq:eom_yielding}
\end{equation}
where the interaction kernel $G_{ij}$ 
(the sum runs over all pair of sites) is chosen to be the Eshelby 
propagator with an amplitude that we control with a factor $k$,

\begin{equation}
G_{ij} \propto k \frac{cos(4\theta_{ij})}{r_{ij}^2}. 
\label{eq:eshelby_real}
\end{equation}

\noindent Notice that this kernel is  long-ranged
(details of the implementation can be found in App.~\ref{app:methods}).
In the case of the amorphous solid, $\Delta X$ corresponds to a change in 
plastic strain ($\Delta X \equiv \Delta\gamma$) instead of interface 
position.
As the depinning counterpart, the yielding transition also displays a thermal 
rounding phenomenon when temperature is relevant~\cite{FerreroPRM2021,PopovicPRE2021}, 
but we stay in the athermal case in the present work. 


The athermal reptation mechanism we are discussing only occurs at driving below 
and sufficiently close to $f_c$.
Yet the effect may be relevant as many systems are expected to adjust spontaneously 
into such a condition.
For instance, the rest slope of a terrain usually accommodates at 
an angle just below the rest angle, as it occurs also with a 
heap of sand or gravel.
This is, the system steps at the situation in which the effect 
of periodic disturbances in the interactions is expected to be maximized.

\section{Results for amorphous solids}
\label{sec:yieldingcreep}

\begin{figure}[t!]
\includegraphics[width=0.95\columnwidth]{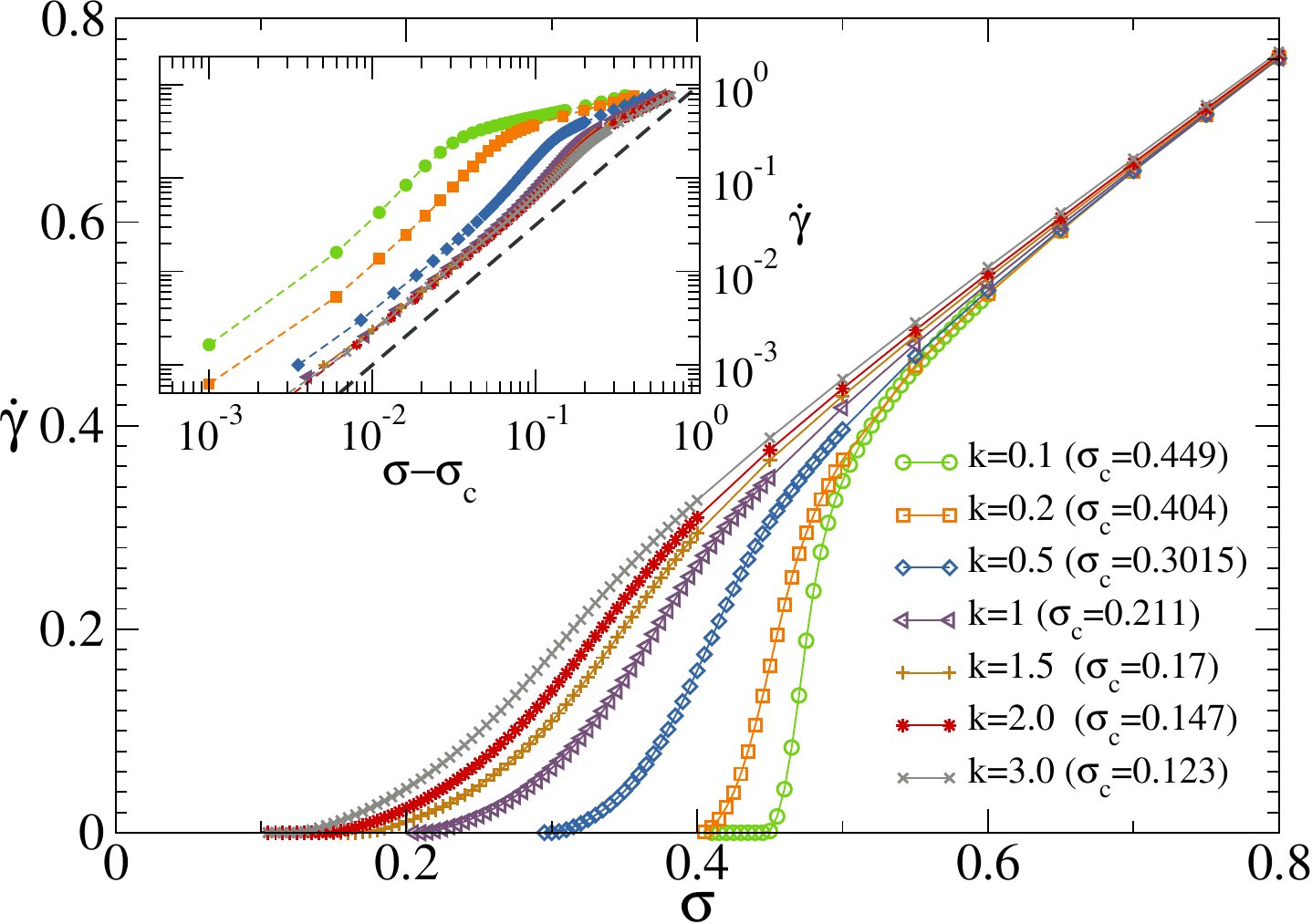}
\caption{
\textit{Yielding model.}
Flowcurves (strain-rate $\dot{\gamma}$ vs stress $\sigma$) for the 
2D elasto-plastic model of an amorphous solid under shear, 
at different values of the long-range elastic 
interaction intensity $k$.
The inset shows $\dot{\gamma}$ vs $\sigma-\sigma_c$, 
where $\sigma_c=\sigma_c(k)$ depends on $k$.
The dashed line displays a law $\dot{\gamma}\sim(\sigma-\sigma_c)^{1.5}$.
Data corresponds to a system of size $N=1024\times1024$.
}
\label{fig:flowcurves_yielding}
\end{figure}

We solve numerically the equations of motion ~\eqref{eq:eom_yielding} 
for different values of the applied stress $\sigma$ and either fix or
oscillating values of $k$. 
Details of the implementations can be found in Appendix~\ref{app:methods}.

First, we characterize the system at fixed values of the elastic 
interaction intensity $k$, by constructing the corresponding 
flowcurves for different values of $k$, Fig.~\ref{fig:flowcurves_yielding}.
This is done starting at a large value of $\sigma$, and calculating the 
average value of $\dot\gamma$ after reaching a steady state along the 
simulation. 
Then, $\sigma$ is progressively reduced and the corresponding values of $\dot\gamma$, 
always in a steady state, are obtained.
We observe in Fig.~\ref{fig:flowcurves_yielding} how the critical force 
increases as $k$ is reduced.
Yet, near $\sigma_c$ all curves behave as $(\sigma-\sigma_c)^\beta$ 
with $\beta\simeq 1.5$, as is shown in Fig.~\ref{fig:flowcurves_yielding}'s inset.
This is a well known results of the yielding transition, and the exponent
is the one expected for the type of disorder potential 
here used~\cite{FernandezAguirrePRE2018, FerreroSM2019}.
It's worth mentioning that, for each value of $k$, the numerical value of 
$\sigma_c$ also depends on the system size $N$.
As is well known, the critical thresholds suffer from finite size 
effects\footnote{Although we haven't performed a systematic finite size analysis
of $\sigma_c$ in this occasion, we have checked that, for a given $N$, the values 
we can infer from a log-lin version of Fig.~\ref{fig:flowcurves_yielding} for 
each $k$ are consistent with the hand-tuned values that give a good power-law 
in the inset.}.

\begin{figure}[t!]
\includegraphics[width=0.95\columnwidth]{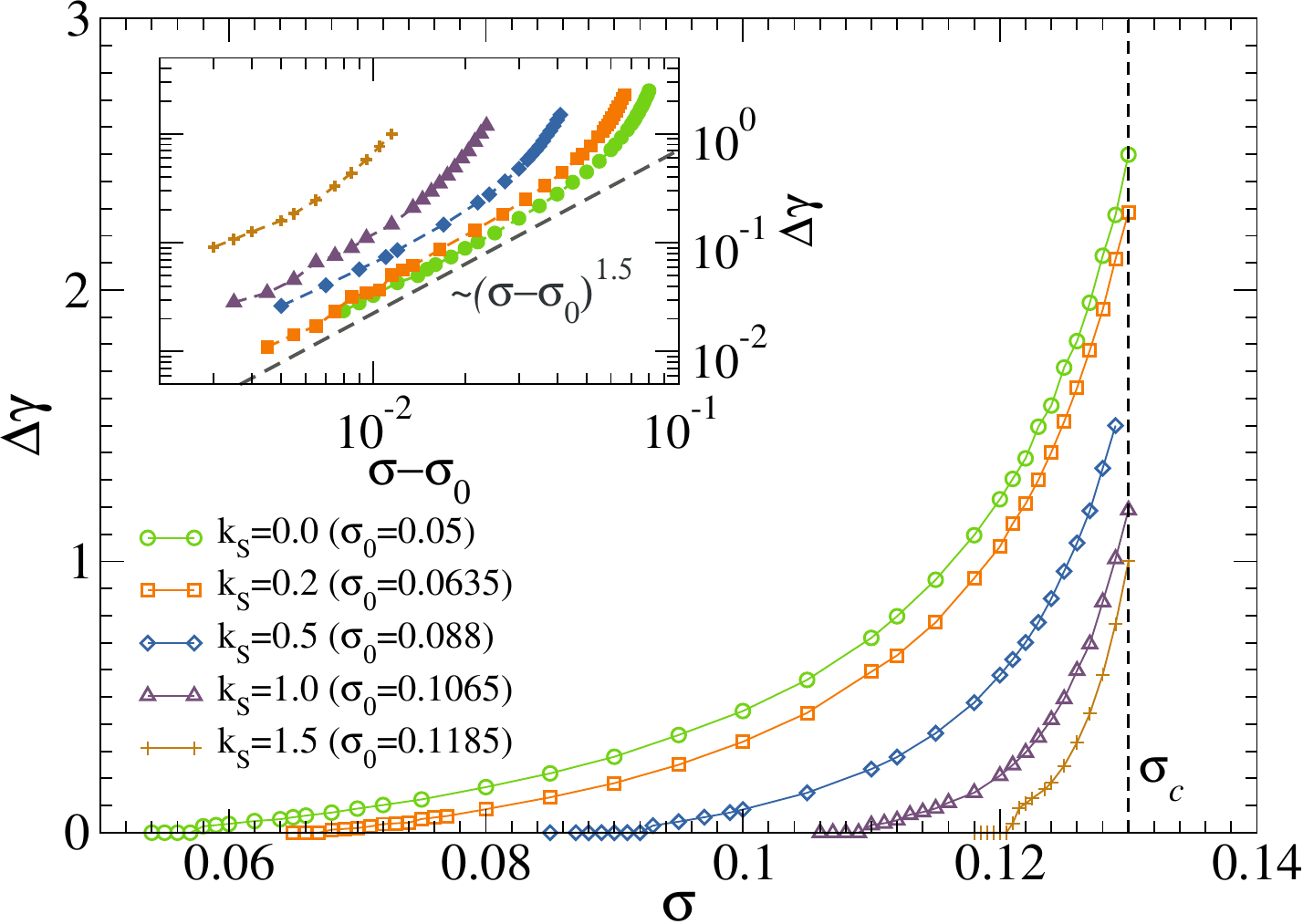}
\caption{
\textit{Yielding model.}
Strain advance $\Delta \gamma$ vs. stress $\sigma$ for our 2D 
elasto-plastic model when cycling between a fixed $k_L=3.0$
and different values of $k_S$. 
The inset displays $\Delta \gamma$ vs. $\sigma-\sigma_0(k_S)$.
The power-law $\sim(\sigma-\sigma_0)^{1.5}$ is displayed as guide 
to the eye.
Data corresponds to a system of size $N=128\times128$.}
\label{fig:Dg_sigma_yielding}
\end{figure}

Now, in the presence of an applied stress below the critical value, 
we cycle the values of $k$ between the starting large value 
$k_L$ and the final small value $k_S$.
This cycling is done very slowly, verifying that a further 
reduction of the cycling rate does not affect substantially 
the results obtained (see Appendix~\ref{app:methods}).
We measure the advance $\Delta\gamma$ of the average strain in 
the system per cycle. 
The results are presented in Fig.~\ref{fig:Dg_sigma_yielding}. 
There is a finite range of stresses starting at $\sigma_c$ and down to
some value $\sigma_0$, in which $\Delta \gamma$ is finite.
For a fix $k_L$, as is the case of Fig.~\ref{fig:Dg_sigma_yielding}, 
the value of $\sigma_0$ depends on $k_S$.
As a matter of fact, the range $\sigma_c$-$\sigma_0$ where oscillations 
produce a non-zero displacement $\Delta\gamma$ becomes wider
as $k_L$-$k_S$ increases.
Note also that, independently of the oscillation amplitude,
$\Delta\gamma$ increases when approaching $\sigma_c$. 

Looking at the inset of Fig.~\ref{fig:Dg_sigma_yielding}, we can further 
point out that the form of $\Delta\sigma$ close to $\sigma_0$ is reminiscent 
of the flowcurves at fix $k$, this is, it reaches $\sigma_0$ with a power-law
consistent with $\Delta\sigma \sim (\sigma-\sigma_0)^{3/2}$.
We will come back to discuss this similarity in Section~\ref{sec:criticality}.

\begin{figure}[b!]
\includegraphics[width=\columnwidth]{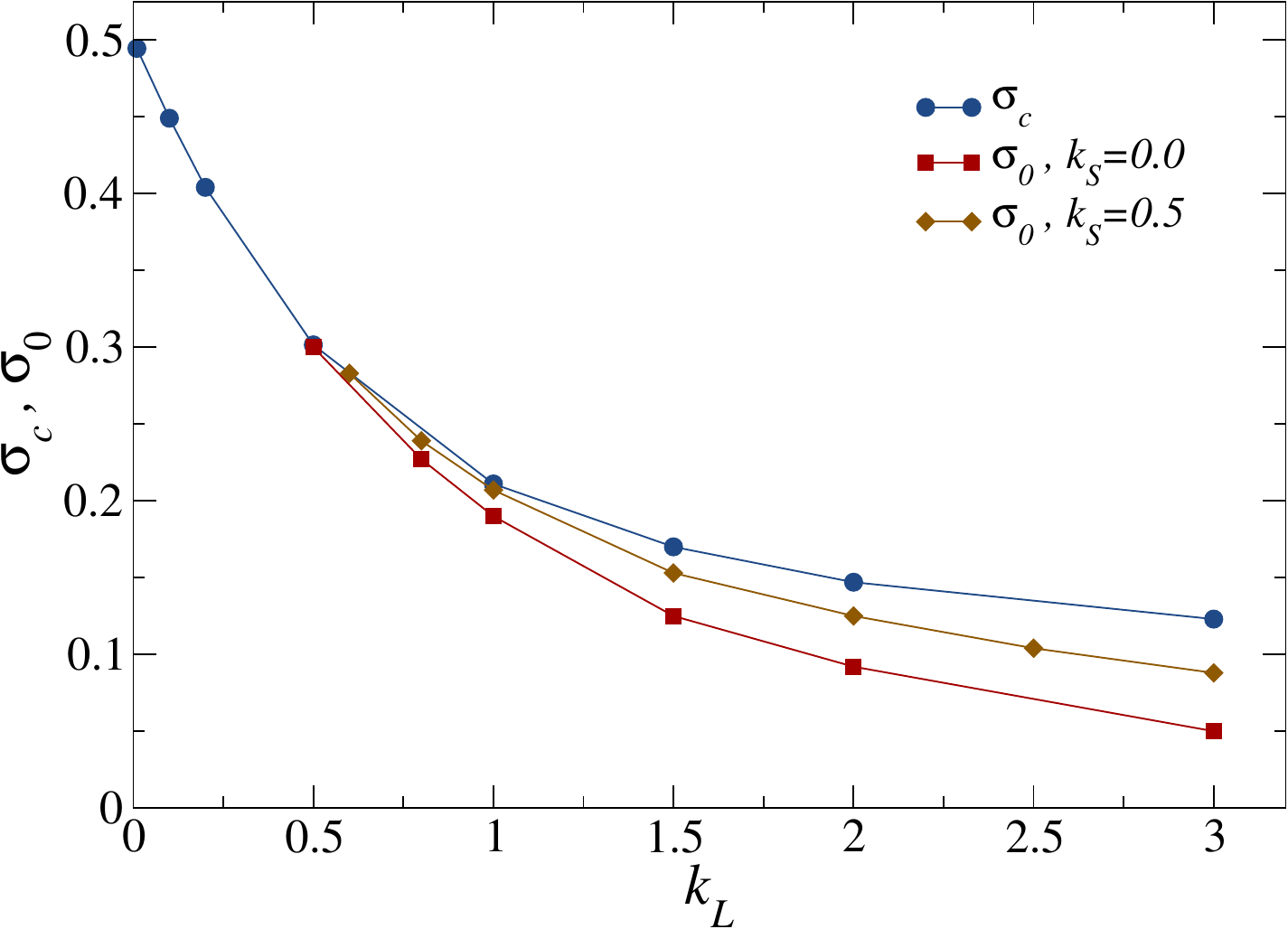}
\caption{
\textit{Yielding model.}
Yielding critical stress $\sigma_c$ (blue circles)
and $\sigma_0$ (red squares for $k_S=0$, chocolate diamonds for $k_S=0.5$) 
vs. $k_L$. 
System sizes used are $N=1024^2$ and $N=128^2$ for the curves of $\sigma_c$
and $\sigma_0$, respectively. 
}
\label{fig:sc_and_s0_vs_kL_yielding}
\end{figure}

The dependence of $\sigma_c$ and $\sigma_0$ with $k_L$ for different values of 
$k_0$ is presented in Fig.~\ref{fig:sc_and_s0_vs_kL_yielding}.
The value of $\eta\equiv (\sigma_c-\sigma_0)/\sigma_c$ is a measurement of the 
relative range in which we observe sub-critical flow. 
We see that $\eta$  is maximal for $k_S=0$, 
and it progressively shrinks as  $k_S$ is increased at a fixed $k_L$.
In addition, for a fixed value of $k_S$  the value of $\eta$ is larger at larger 
values of $k_L$, and it decreases as $k_L$ does. 
The mean field analysis of Sec.~\ref{sec:meanfieldcreep} suggests that $\eta$ is 
different from zero in all the range $k_L>k_S$, yet it is very small when $k_L\gtrsim k_S$.

\section{Results for elastic surfaces}
\label{sec:depinningcreep}

\begin{figure}[t!]
\includegraphics[width=\columnwidth]{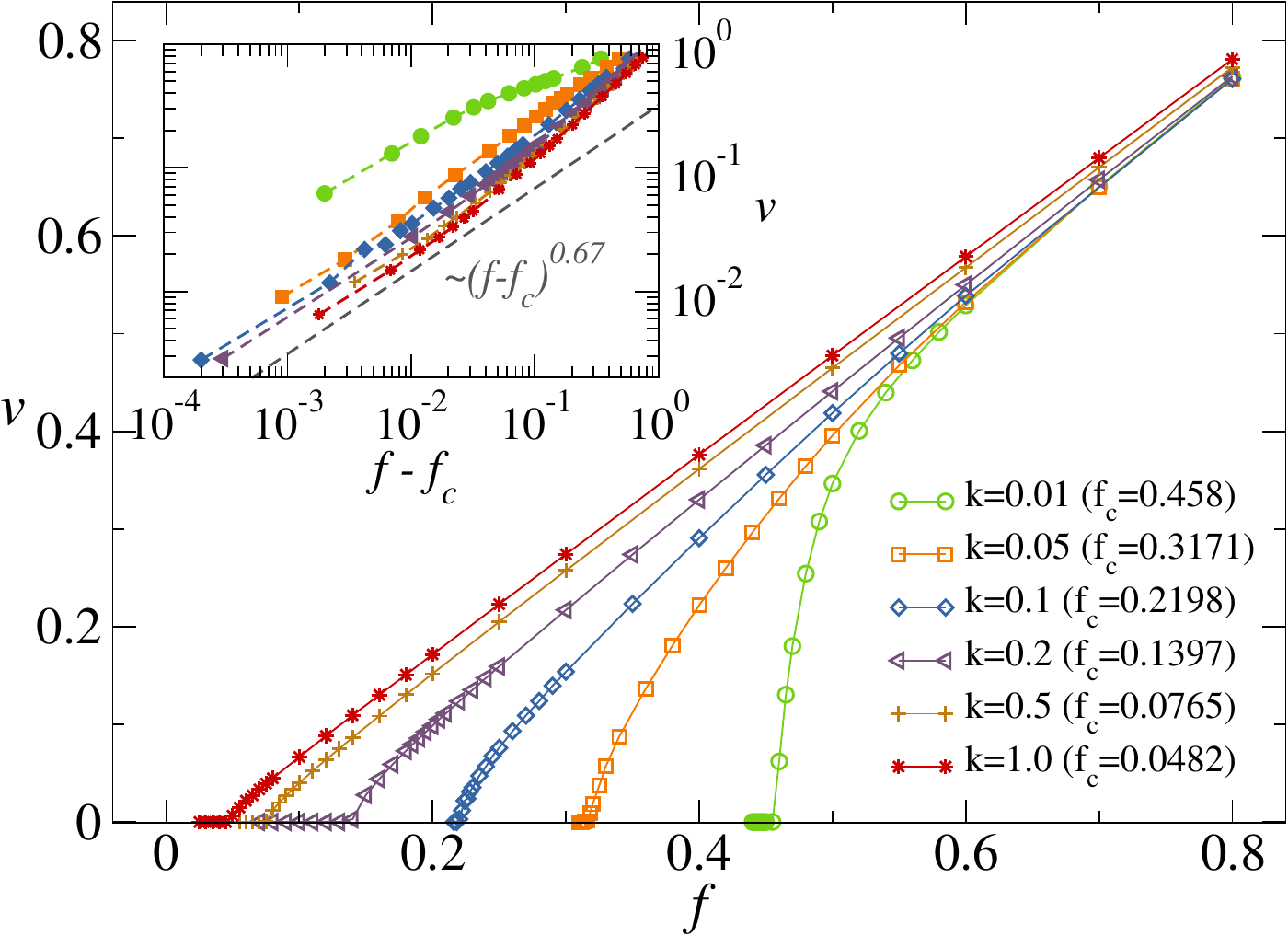}
\caption{
\textit{Depinning model.}
Velocity-force characteristics ($v$ vs $f$) for the 
2D elastic interface model of depinning, 
at different values of the short-range elastic 
interaction constant $k$.
The inset shows $v$ vs $f-f_c$, where $f_c=f_c(k)$ depends on $k$.
The dashed line displays a law $v\sim(f-f_c)^{0.67}$.
Data corresponds to a system of size $N=128\times128$.
}
\label{fig:flowcurves_depinning}
\end{figure}

We now present results for the effects of changing or oscillating
the elastic interaction strength in the case of elastic interfaces 
on disordered media that undergo a depinning transition.
We proceed in analogy with the yielding model case of the previous section,
but now solving Eq.~\ref{eq:eom_depinning}, either with fixed, or oscillating $k$ values.

First, we characterize the system at fixed values of $k$, by 
constructing the corresponding flowcurves, as displayed in  Fig.~\ref{fig:flowcurves_depinning}. 
Note how the critical force $f_c$ increases as $k$ is reduced.
Still, near $f_c$ all curves behave as $(f-f_c)^\beta$ with 
$\beta\simeq 0.67$, as is shown in the inset.
A  value of $\beta$ smaller than 1 is a well known behavior of depinning 
models, and it is an important difference with the yielding case, where 
$\beta>1$.
Moreover, we observe compatibility with the exponent expected for 
short-range depinning in $d=2$~\cite{Leschhorn1993, FerreroARCMP2021}.

\begin{figure}[t!]
\includegraphics[width=0.98\columnwidth]{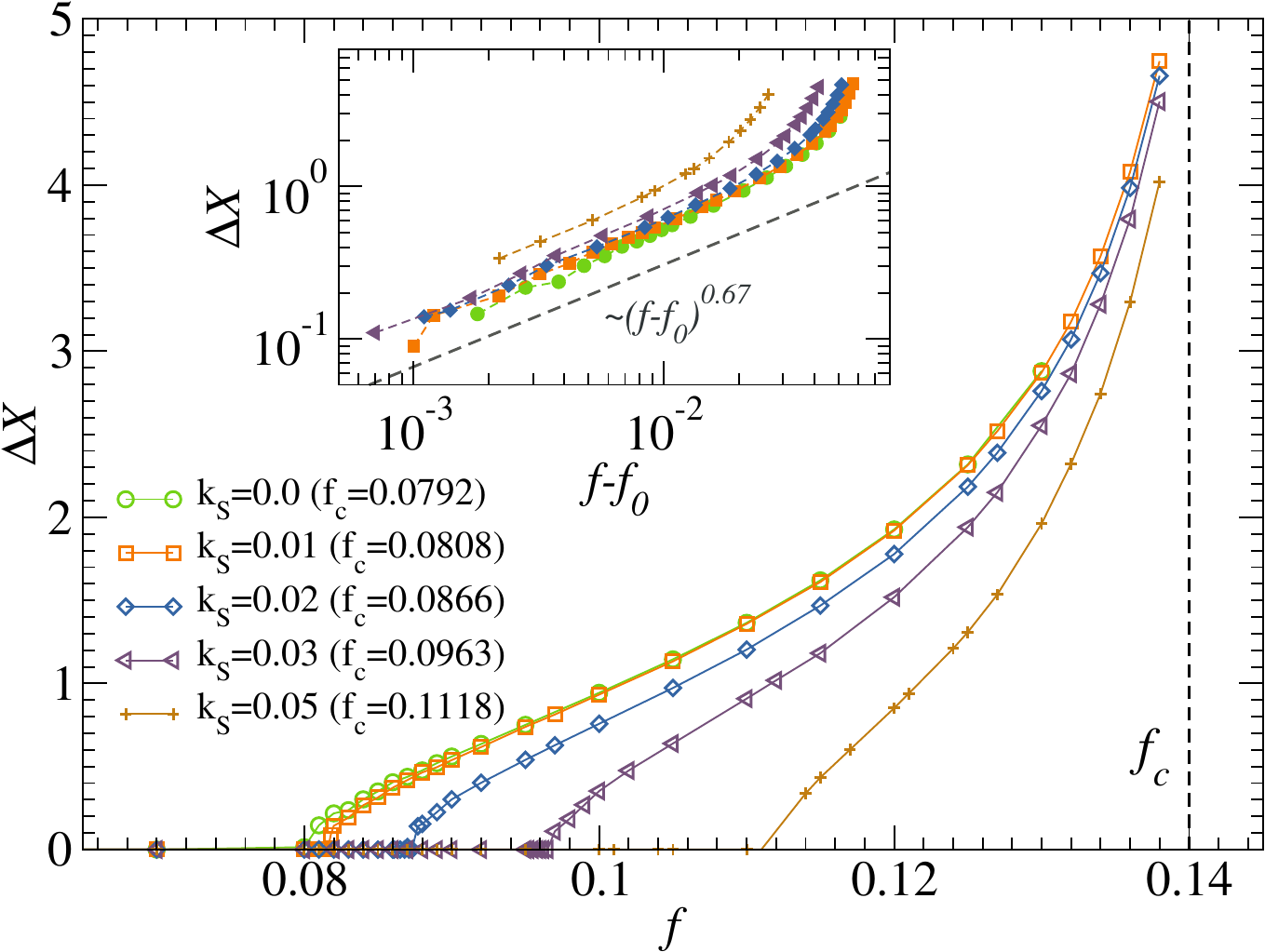}
\caption{
\textit{Depinning model.}
Interface advance $\Delta X$ vs (sub-critical) applied force $f$ 
when cycling between a fixed $k_L=0.2$ and different values of $k_S$.
The inset displays $\Delta X$ vs. $f-f_c(k_S)$.
The power-law $\sim(f-f_c)^{2/3}$ is displayed as guide to the eye.
Data corresponds to a system of size $N=128\times128$.
}
\label{fig:DX_vs_f_depinning_varying_ks}
\end{figure}

Now, in the presence of an applied force smaller than the critical one, 
we cycle the values of $k$ between the starting value $k_L$ and the 
a final value $k_S$.
As in the previous case, we observe an advance $\Delta X$ per cycle
of the interface; results are presented in 
Fig.~\ref{fig:DX_vs_f_depinning_varying_ks}. 
There is a finite range of forces between some $f_0$ and the critical force
$f_c$,  in which the value of $\Delta X$ is finite. 
Note how $\Delta X$ increases when approaching $f_c$. 
The range $[f_0,f_c]$ where the effect is observed becomes wider when the 
separation $k_L$-$k_S$ increases.
Since we work in Fig.~\ref{fig:DX_vs_f_depinning_varying_ks} at a 
fixed $k_L=0.2$, the range of forces at which the subcritical athermal
reptation occurs is maximal for $k_S=0$.
Again, as in the yielding case, we point out that the form of $\Delta X$ 
close to $f_0$ seems to be consistent with a `shift' of the criticality 
from $f_c$ to $f_0$, i.e., $\Delta X$ maintains the $\beta\simeq 2/3$ 
exponent of the velocity-force characteristics around $f_c$ on its 
behavior close to $f_0$: $\Delta X \propto (f-f_0)^{2/3}$.

\begin{figure}[b!]
\includegraphics[width=\columnwidth]{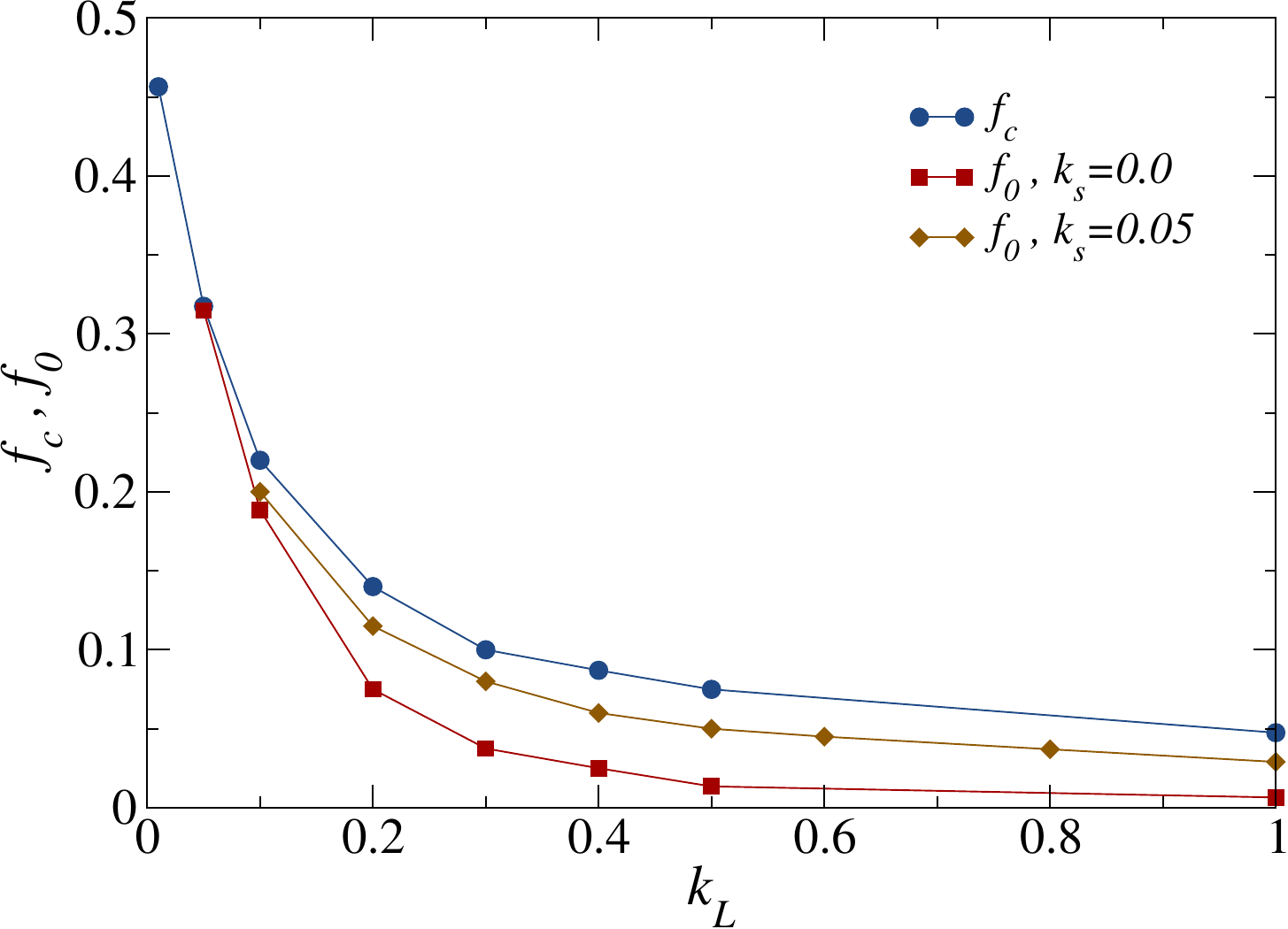}
\caption{
\textit{Depinning model.}
Critical force $f_c$ (blue circles)
and $f_0$ (red squares for $k_S=0$, chocolate diamonds for $k_S=0.05$)
vs. $k_L$. 
System size used is $N=128^2$ 
}
\label{fig:fc_and_f0_vs_k_depinning}
\end{figure}

The dependence of $f_c$ and $f_0(k_S)$ with $k_L$ is presented 
in Fig.~\ref{fig:fc_and_f0_vs_k_depinning}.
Both $f_c$ and $f_0$ decrease as $k$ (or $k=k_L$) is increased,
nevertheless, $f_0$ drops faster, specially when the oscillation
amplitude $(k_L-k_S)$ is large.
The difference between $f_c$ and $f_0$ for a given $k_L$ 
allows for a window of observation of finite advances 
$\Delta X$ of the interface through the mechanisms of 
athermal reptation  facilitated by oscillations of $k$. 
As in the yielding case, for any fixed value of $k_S$
the value of $\eta\equiv (f_c-f_0)/f_c$ decreases as $k_L$ does.
Again, the mean field results in Sec.~\ref{sec:meanfieldcreep}  suggest
that $\eta$ only vanishes in the limit $k_S\to k_L$.

\section{Criticality  at $f_0$}
\label{sec:criticality}

\begin{figure}[t!]
\includegraphics[width=\columnwidth]{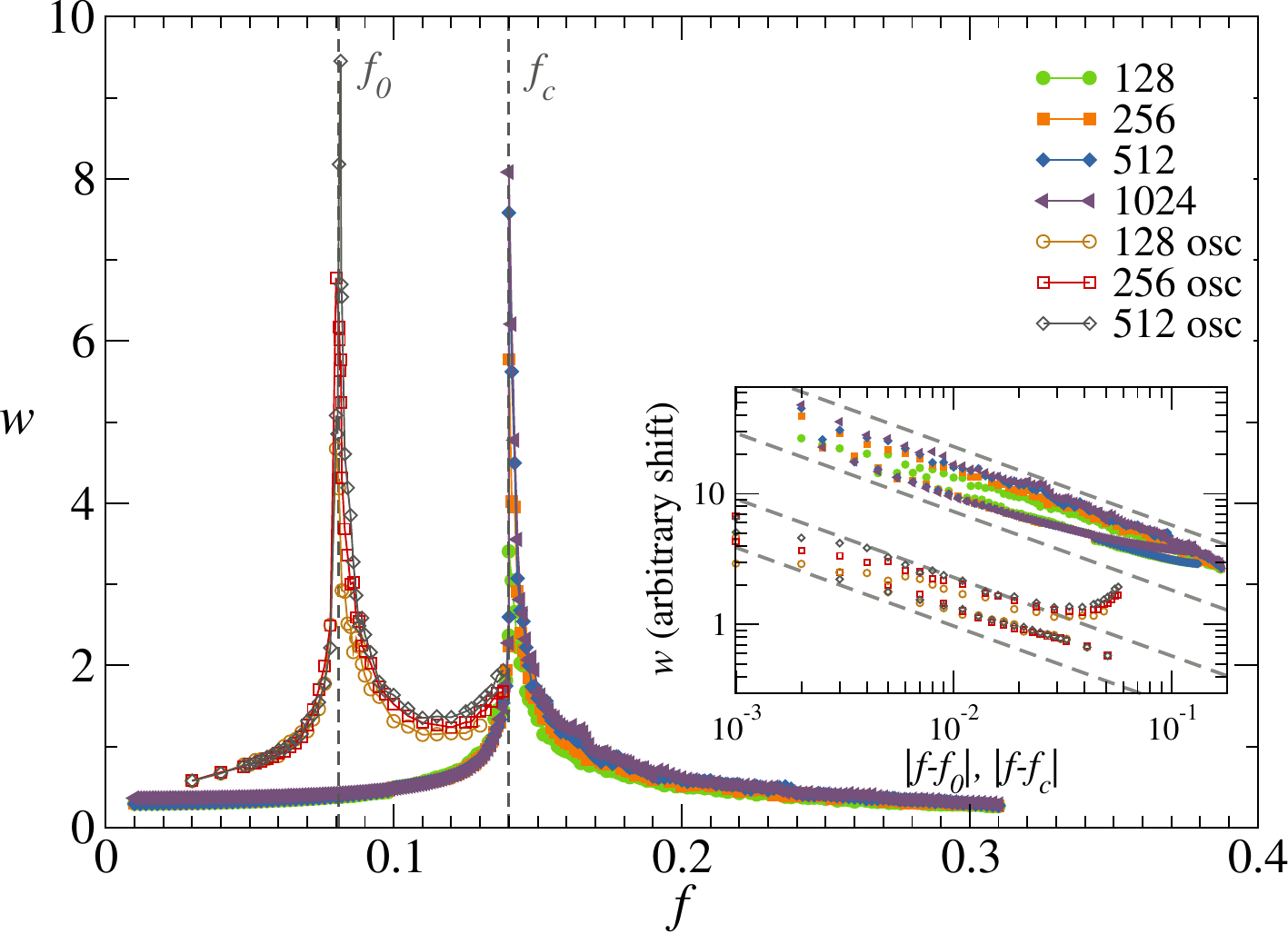}
\caption{
\textit{Depinning model.}
Interface width $w$ as a function of force for both constant $k=k_L=0.2$
and cyclic $k$ (between $k_L$ and $k_S=0$), and different system sizes.
Vertical dashed lines mark the location of $f_c$ and $f_0$.
The inset shows $w$ vs $|f-f_c|$ or $|f-f_0|$ in log-log scale
(both `from-the-right' and `from-the-left' branches are shown for 
each peak, and curves corresponding to the peak at $f=f_c$ are shifted 
up a factor of ten for a better visualization).
The gray dashed lines there correspond to $w~\sim |f-f_c|^{0.6}$.
}
\label{fig:w_depinning}
\end{figure}

The form of the flowcurves above and close to the depinning critical 
value $f_c$ ($\sigma_c$ for yielding) is characterized by an 
exponent $\beta$, which contains information about the criticality 
of the depinning (or yielding) transition.
\begin{equation}
 v \sim (f-f_c)^{\beta_d} \quad ; \quad \dot{\gamma} \sim (\sigma-\sigma_c)^{\beta_y}
\end{equation}
We have seen that, when analyzing the curves depicted by the sub-critical 
advance \textit{per cycle} $\Delta X$ (by oscillations in $k$) close to $f_0$, 
they also look as power-laws.
Moreover, the exponents $\beta$ ($\beta_d \sim 2/3$ for depinning and 
$\beta_y \sim 3/2$ for yielding) seemed to be conserved (within the precision of
our numerical data), namely
\begin{equation}
 \Delta X \sim (f-f_0)^{\beta_d} \quad ; \quad \Delta\gamma \sim (\sigma-\sigma_0)^{\beta_y}
\end{equation}

This similarity raises the question about the possibility of having a criticality 
analogous to the one of the parent transition (depinning/yielding) 
but at $f_0$ ($\sigma_0$) in the problem of sub-critical advance with 
oscillations in $k$.
We present now further evidence of criticality around
$f_0$ ($\sigma_0$), which favors the hypothesis that the parent transition
at $f_c$ ($\sigma_c$) is translated somehow to the new (lower) thresholds when 
oscillations in the environmental conditions step in.

For depinning with constant $k$ it is well known that, 
in analogy with equilibrium critical phenomena, there is a correlation 
length $\xi$ diverging at $f_c$ as ${\xi\sim(f-f_c)^{-\nu}}$ in the thermodynamic 
limit, which is the hallmark of criticality in the system~\cite{FisherPR1998, Kardar1998}.
One way to assess this correlation length 
is to evaluate the interface width $w$ as a function of $f$,
which close to $f_c$ is expected to scale as $w\sim\xi^{d+\zeta}$,
with $\zeta$ the roughness exponent~\cite{FerreroCRP2013, FerreroARCMP2021}. 
We may ask if a similar divergence exists in the case of subcritical 
interface advance under cycling of $k$, but now around $f=f_0$.  
Using the standard definition of width 
\begin{equation}
 w^2\equiv {\overline {x_i^2}-\overline x_i^2},    
\end{equation}\label{eq:width}
\noindent we investigate the value of $w$ as a function of $f$,
in finite systems of different sizes and 
comparing the cases of fixed and cycled values of $k$.

\begin{figure}[t!]
\includegraphics[width=1\columnwidth]{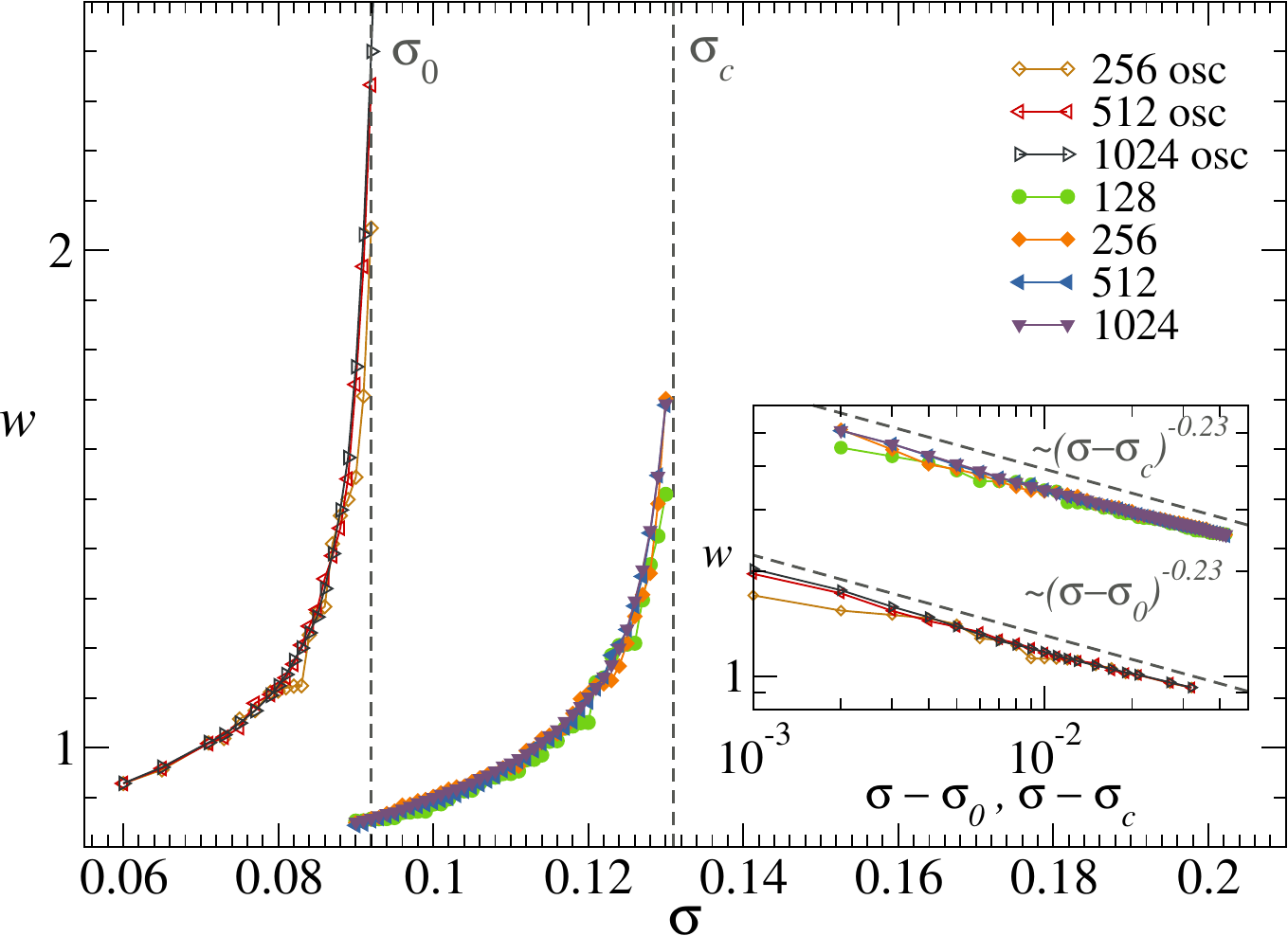}
\caption{
\textit{Yielding model.}
Yielding strain manifold width $w$ vs $\sigma$ for both constant $k=k_L=3.0$
and cyclic $k$ (between $k_L$ and $k_S=0.5$), and different system sizes.
Vertical dashed lines mark the location of $\sigma_c$ and $\sigma_0$.
The inset shows $w$ vs $\sigma-\sigma_c$ or $\sigma-\sigma_0$ in log-log scale
and the gray dashed lines there correspond to $w~\sim (\sigma-\sigma_{c,0})^{0.23}$
(curves corresponding to the peak at $\sigma=\sigma_c$ are shifted up a factor of 
three for a better visualization).
}
\label{fig:w_yielding}
\end{figure}

In the case of a constant $k$ we proceed in the following way. 
We start at $f=0$ with a flat interface ($x_i=0$) and allow 
the interface to adapt to the pinning forces.
Then, $f$ is increased slowly, allowing for the interface to reach 
a stable configuration and calculating the value of $w$ at each $f$. 
A stationary situation with progressively larger $w$ is reached only 
if $f<f_c$.
When $f>f_c$ the interface is continuously evolving (advancing)
but this does not introduce any problems in the calculation of $w$,
which decreases as $f$ departs from $f_c$ to larger and larger 
values~\footnote{Yet, reaching the true steady state for each
value of $f$ is not equally easy when coming from bellow $f_c$
increasing $f$ and when starting from $f\gg f_c$ and decreasing
it. We check consistency by simulating both protocols and with 
this we estimate an uncertainty in $f_c$.
}.
In the end, we obtained the curves shown in Fig.~\ref{fig:w_depinning}. 
A sharp maximum of $w$ around $f_c$ is clearly observed, as a 
sign of criticality. 
Consistently with what is expected, we see that the maximum of $w(f)$ 
increases with system size. 
Moreover, we can study how does $w(f)$ behaves around $f=f_c$.
For depinning one expects~\cite{FerreroARCMP2021} 
$w\sim \xi^\zeta$ and $\xi\sim(f-f_c)^{-\nu}$, therefore
$w\sim (f-f_c)^{-\nu\zeta}$, which in the case of $d=2$ short-range
depinning ($\zeta=0.8$, $\nu=0.8$) is $w\sim (f-f_c)^{-0.6}$.

Then, we perform a similar analysis under cycling $k$ between 
$k_L$ and $k_S$. 
The protocol is unchanged with respect to the constant $k$ case,
with the important clarification that values of $w$ are now taken 
stroboscopically in the moments when $k=k_L$~\footnote{Let us also 
mention that in order to achieve convergence in the 
values of $w$ obtained, a relatively large number of cycles in $k$ 
must be performed at each value of $f$.
}
Values of the width $w$ in the oscillatory regime are shown 
in Fig.~\ref{fig:w_depinning} alongside those obtained  at fix $k$. 
We clearly observe a peak of the interface width at $f_0$, that separates the 
regions of no cyclic advance ($f<f_0$) from that of cyclic interface advance ($f>f_0$).
This peak suggests that we are in the presence of a critical configuration of 
the interface at $f_0$ when cycling $k$, analogous to the critical configuration 
at $f_c$ under constant $k$.

Switching now to the case of yielding, let us start by recalling some inherent problems 
with the definition of the interface width $w$ in such case.
As it is well known, the Eshelby kernel possesses soft modes~\cite{Cao2018} 
(i.e., directions in ${\bf q}$ space with vanishingly small energy) that are responsible 
for an unbounded increase of the interface width in time, when the interface is moving.
This causes the value of $w$ to be ill defined, since it typically increases in a 
diffusive way with time.
However, this occurs only in the moving phase (i.e., for $\sigma>\sigma_c$ in the 
constant $k$ case, or $\sigma>\sigma_0$ in the cycling case), whereas the value 
of $w$ can still be defined below the critical values $\sigma_c$ or $\sigma_0$. 
Therefore, for yielding  we present results in those regions of applied stresses.
Fig.~\ref{fig:w_yielding} shows the results obtained for the width $w$
of the elastic manifold in the elastoplastic model simulations 
(using the same definition as for depinning, eq.~\ref{eq:width}).
We see a divergence of $w$ close to $\sigma_c$ for the constant $k$,  
and a similar one close to $\sigma_0$ for the oscillating $k$ situation
of the kind $w\sim(\sigma-\sigma_{c,0})^{0.23}$.
\footnote{
For yielding in $d=2$ dimensions, estimations of $\nu$ exist 
(e.g. $\nu\simeq 1.16$ ~\cite{LinPNAS2014}); 
so our result can be used to guess -through scaling arguments- 
the `roughness' of the elastoplastic manifold at criticality. We
leave this discussion  for a future work.}
As in the depinning case, this suggests a critical configuration of the
interfaces at $\sigma_0$ under oscillation of $k$, similar to that 
occurring at  $\sigma_c$ under constant $k$.

\section{The reptation mechanism in a mean-field approach}
\label{sec:meanfieldcreep}

The results in the previous sections concerning spatially distributed 
depinning and yielding models are the closest to 
geophysical application and can serve as a starting point for more 
realistic studies.
Nevertheless, we think it is conceptually valuable  to complement 
those results with a mean-field approximation.
This will give insight into the mechanism of sub-critical deformation, and will 
also allow us to analytically verify some of the claims that we made in the 
presentation of results for spatially extended models.

Let us consider a system of $N$ particles characterized by their coordinates 
$x_i$ ($i=1,...,N$) interacting elastically. 
The mean field nature of the model is contained in the form of the elastic 
interaction, that produces an elastic force on each particle given by
\begin{equation}
f_i^{el}=k(X-x_i)    
\end{equation}
where $X=N^{-1}\sum_i x_i$ is the average position of the interface.
Furthermore, in the present section we take the potential $V_i(x_i)$ of 
interaction with the substrate to be a collection of narrow wells randomly 
distributed along the $x_i$ coordinate with a mean separation $a$. 
This can be thought to correspond to a limit in which the parabolic wells used 
previously become very narrow.
The wells are characterized by the maximum force $f_p$ that needs 
to be applied to a particle trapped in the well to escape from it. 
For simplicity, we take the value of $f_p$ to be the same for all wells,
stochasticity is guaranteed by the random position of the wells.
In addition, an external force $f$ is assumed to be applied to the particles. 

In the narrow well approximation the dynamical evolution equation 
(of the kind of Eq.~\ref{eq:eom_depinning}) is replaced by a discrete 
rule, defined in the following way. 
If a particle is inside a potential well, it remains there as long as
the  absolute value of the force on the particle $F_i\equiv f+f_i^{el}$ 
is lower than the pinning force $f_p$. 
If $|F_i|>f_p$, in a single time step the particle jumps (towards the right 
or the left according to the sign of $F_i$) to the equilibrium point where $F_i=0$, 
namely $\widetilde {x_i}=f/k+X$, or to a new potential well if it happens to reach 
one in between $x_i$ and $\widetilde {x_i}$. 

The critical force $f_c$ in this model is the maximum value of $f$ for 
which a stationary (non-moving) situation can be found, namely a configuration 
in which all sites have either $|F_i|<f_p$ and are within pinning centers, 
or have $F_i=0$.
The value of $f_c$ can be obtained analytically  
(see Appendix~\ref{app:analyticMF}).
Introducing the rescaled variable
\begin{equation}
z =\frac{ka}{f_p}
\end{equation}
one obtains that $f_c$ is given by
\begin{equation}
f_c=f_p\left (1-z+ze^{-1/z}\right )\label{sigmac}.
\end{equation}

\begin{figure}[t!]
\includegraphics[width=0.5\columnwidth]{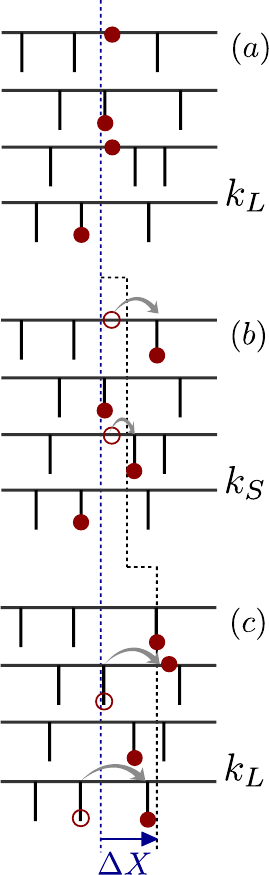}
\caption{Qualitative evolution of the configuration of the system when $k$ passes 
from a high value $k_L$ (a) to a small one $k_S$ (b) and increases again to $k_L$ (c). 
The same four spatial positions $i$ are depiced in each case.
A constant stress $f$ (pointing to the right) is present in all cases. 
Although the parameters of the system in (a) and (c) are identical, the 
system configuration is not, the mean position moved to the right from (a) to (c). 
}
\label{fig:scatter2}
\end{figure}

Now, we introduce in the model the variation in time of the spring constant $k$, 
considering a cyclic variation between a large value $k_L$, and a small value $k_S$, 
and take this variation to occur quasi-statically, this is, not introducing effects 
associated to the velocity of variation.
The process can be analyzed qualitatively as follows (refer to Fig.~\ref{fig:scatter2}).
We suppose that the system is under an applied force $f$ that is lower 
than $f_c$ for all values of $k$ in the range $k_S$-$k_L$ (in practice, 
this means that $f$ is lower than the $f_c$ corresponding to $k_L$). 
In Fig.~\ref{fig:scatter2}(a) we sketch a configuration of the system at a 
large value $k_L$ of $k$. 
This is a stable configuration, with some particles at pinning centers, 
and some others outside them.
In Fig.~\ref{fig:scatter2}(b) we depict the configuration of the system when 
$k$ has been reduced to a value $k_S$ that for a simpler analysis has been 
taken to be zero. 
Sites that were pinned in (a) remain pinned at the same well, but those that 
were unpinned are dragged to the right by $f$, and each one reaches the first 
available well, where it gets pinned.
In Fig.~\ref{fig:scatter2}(c) the value of $k$ is increased again to $k_L$ 
and some particles (those located in the left-most wells) jump out of their 
pinning centers, as the total force on them is larger than $f_p$.
The system accommodates in a new equilibrium configuration (c) that is not
coincident with the one in (a), although the parameters in (c) are the same 
as those in (a).
Therefore, there is a finite shift in the mean position of the interface 
$\Delta X\equiv X_c-X_a$. 
If the cycling of $k$ between $k_L$ and $k_S$ is repeated, a shift $\Delta X$ 
is expected to occur on each cycle.  
The value of $\Delta X$ will be larger when $f$ is close to $f_c$ and 
will be smaller as $f$ is decreased away from $f_c$.
This is expected, since $f$ is the driving force for the increase of $X$ 
on each cycle of variation of $k$.

\begin{figure}[t!]
\includegraphics[width=\columnwidth]{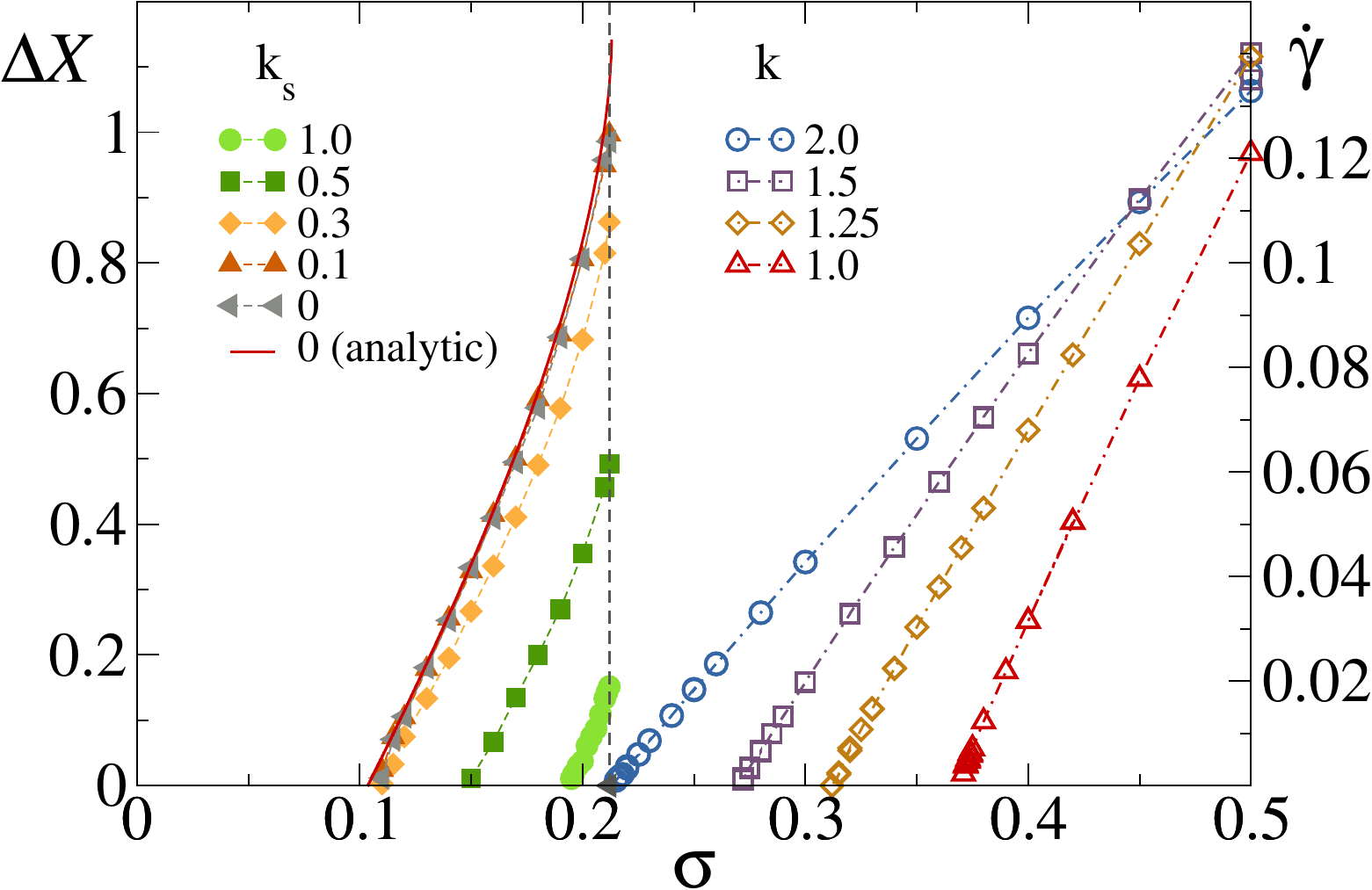}
\caption{
\textit{Fully-connected depinning model.}
(right) Numerical flow curves of the system at different values of $k$. 
(left) Interface advance $\Delta X$ per cycle, as a function of applied force $f$ 
when the interface stiffness is cycled between $k_L=2$, and values of $k_S$ as indicated.
Points are the results of numerical simulations. The continuous red line is the analytical result
for $k_S=0$ (See App.~\ref{app:analyticMF}). 
}
\label{fig:delta_x_medio}
\end{figure}

By the treatment presented in App.~\ref{app:analyticMF} we have been able 
to derive analytically the form of $\Delta X$ as a function of $f$ 
($f<f_c$) in the case in which $k_S=0$. 
This is shown in Fig.~\ref{fig:delta_x_medio} with the continuous red line. 
This analytical result is very important as it shows that there is in fact a 
minimum value $f_0$ that has to be exceeded to have a finite value of $\Delta X$.
Also, note that in the present case the value of $\Delta X$ when $f\to f_c$ is 
finite, but it has an infinite derivative at that point.
For $k_S=0$ the analytical expression we obtain for $f_0$ is
\begin{equation}
f_0=f_p\left(1-z+(1+z)e^{-2/z}\right)    
\end{equation}

\begin{figure}[b!]
\includegraphics[width=0.95\columnwidth]{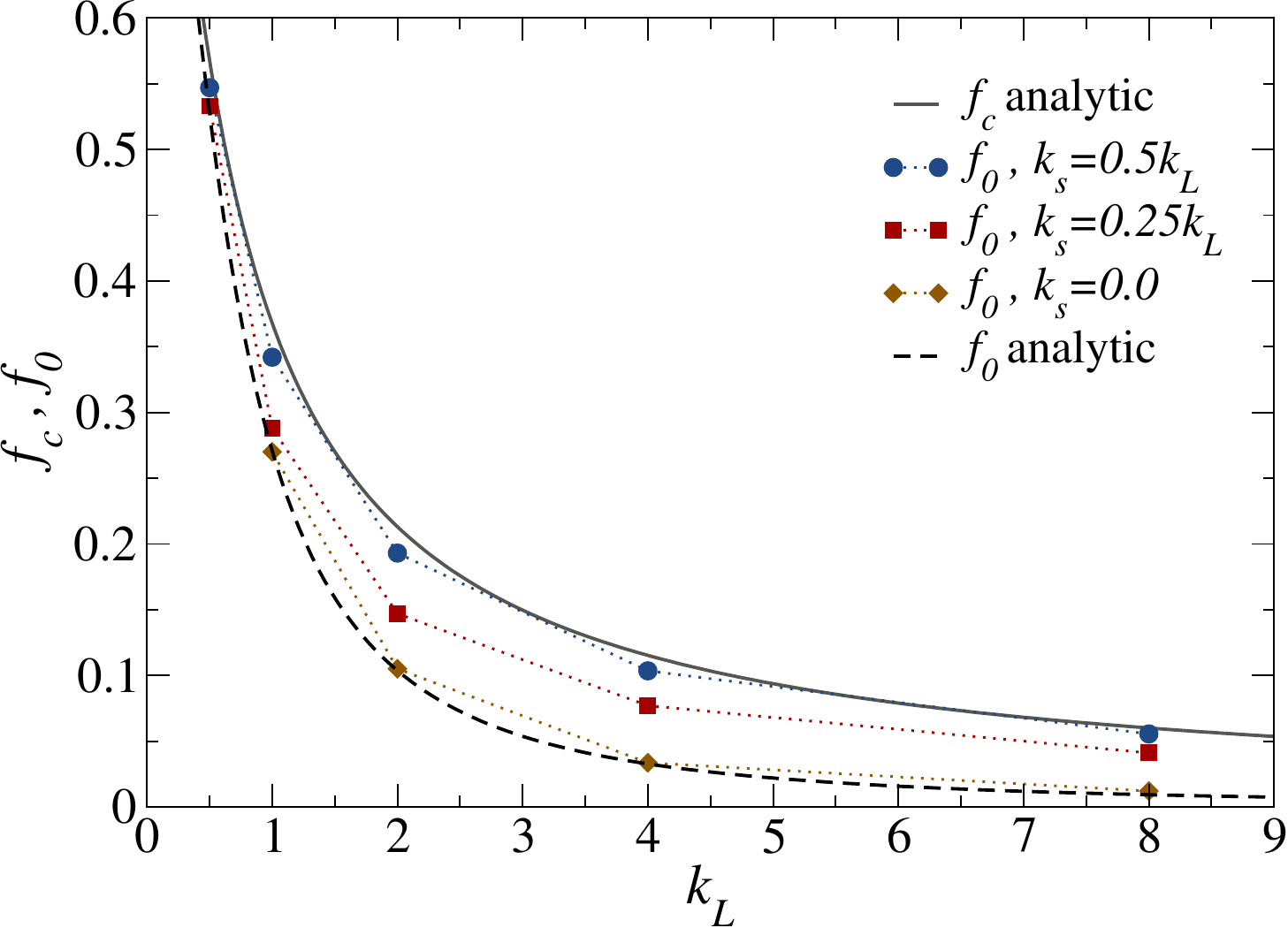}
\caption{
\textit{Fully-connected depinning model.}
Critical force $f_c$ as a function of $k_L$ (gray line, analytical result),
reptation limit $f_0$ as a function of $k_L$ (black-dashed line, analytical result),
and numerical results for $f_0$ when cycling the system between 
$k_L$ and $k_S$ as indicated in the labels. 
}
\label{fig:s1s2}
\end{figure}

This expression, together with Eq.~\eqref{sigmac} for $f_c$ are plotted 
in Fig.~\ref{fig:s1s2}. 
Although the two curves become very close as $k_L$ approaches zero, 
they remain different for any $k_L\ne 0$. 
We believe this also occurs in the depinning and yielding 
cases (Figs. (\ref{fig:sc_and_s0_vs_kL_yielding}) and 
(\ref{fig:fc_and_f0_vs_k_depinning})); 
notice that curves there show data obtained at fixed values of 
$k_S$ and not relative to $k_L$,
therefore the behavior is not as evident as in Fig.~\ref{fig:s1s2}.

We complement the analytical results with numerical simulations to 
obtain the value of $\Delta X$ as a function of the values $k_L$ 
and $k_S$, at different values of the applied force $f$.
A system with $N=10^5$ sites is simulated following the rules 
explained at the beginning of the section. 
First, a value of $k=k_L$ is chosen and some $f>f_c$ is applied 
during a number of steps to obtain a steady state.
Then we reduce progressively $f$ repeating the procedure 
and measuring in the steady states to obtain the flow curve.
When $f$ becomes lower than $f_c$, $X$ sets to a constant value.
Starting from this initial configuration we slowly cycle $k$ between
$k_L$ and the chosen value of $k_S$ and obtain the average advance 
of the interface $\Delta X$ per cycle. 

When $f>f_c$, the deformation of the system increases at a finite 
rate with time, defining the flow curve $\dot X$ vs $f$.
This is plotted in the right part of Fig.~\ref{fig:delta_x_medio} for 
different values of the parameter $k$. 
As in the spatially extended models for depinning and yielding of previous sections, 
we see how lower values of $k$ displace the curves to the right:
softer elastic interaction gives possibility to the system to accommodate 
better to the pinning potential and the necessary stress $f_c$ to produce 
a finite deformation velocity increases.
The results for $\Delta X$ are displayed in the left part of 
Fig.~\ref{fig:delta_x_medio}, where $\Delta X$ is shown as a function of $f$, 
when $k$ is cycled between a value $k_L=2$, and different values of $k_S$. 
For $k_S=0$ the numerical results nicely reproduce the analytical ones. 
When $k_S>0$ the range $f_0$-$f_c$ in which the effect is observed is reduced. 

In Fig.~\ref{fig:s1s2} we also show results of numerical simulations for $f_0$ and $f_c$.
We see that numerical and analytical values of $f_c$ as a function of $k=k_L$ agree very well.
The same occurs for $f_0$ when $k_S=0$.
The region between $f_0$ and $f_c$ is the range in which there is a non-zero advance
$\Delta X$ in each cycle of variation of $k$.

\section{Discussion and connection with related phenomena}

Let us now emphasize similarities and differences between the mechanism 
of sub-critical flow or reptation presented here and other cases that have 
been considered previously in the literature.
As stated in the introduction, external mechanical noise has been studied 
as a possible driver of sub-critical flow in soft-glassy materials.
In many cases, external noise is assumed to act randomly in time and/or 
space, making its effect similar to that of thermal noise, apart from 
differences in relative intensity~\cite{BontempsNC2020, AuzetESPL1996,GabetESPL2000}. 
In other cases, the external perturbation acts rather homogeneously 
across the system, as for instance in the case of 
cyclic loading~\cite{GarciaRojoGM2005, YuanNatComm2024},
or  when a ``tapping"  noise is applied~\cite{DeshpandeNC2021}. 
In addition, ``noise" has been applied on top of an average external 
stress as a stochastic contribution~\cite{LeGoffPRL2019}, or simply as 
a serrated contribution to the stress~\cite{PonsPRE2015}. 
In this last case a viscous response directly related to small stress 
modulations and consequently `flow' below the yield stress is found, 
in a scenario described as a `secular drift' or ratcheting process 
at long times.

The key distinction between previous scenarios and our results 
is that we consider a perturbation (the variation in the elastic 
stiffness $k$ of the system) that acts in a quasi-static limit, 
meaning it has no effects associated with its rate of change.
In addition, this perturbation is homogeneously applied to the whole 
system, and it can be considered to be rooted in variation of 
environmental conditions.
We have described our mechanism as a ``reptation" process, which is 
an image particularly adapted to the two-particle model of Section~\ref{sec:twoparticles}, 
as well as for the model system described by Moseley~\cite{MoseleyPRSL1856}. 
Yet, the full models of depinning and yielding that we considered 
can be qualitatively described by the same basic mechanism.

It is worth commenting on the literature on
thermally cycled granular systems~\cite{DivouxPRL2008, PercierEPL2013, 
PastenAG2019, CoulibalyGM2020, RottaLoriaGEE2021, PanGM2024, PanSR2024}.
Some experimental setups incorporating thermal cycling are 
related to pile compaction~\cite{DivouxPRL2008,PanSR2024}; while 
others by including a lateral forcing on a body resting on 
the granular system~\cite{PastenAG2019} analyze ratcheting displacement, 
and could bare more similarity with the downhill soil creep.
Irrespectively, in all these cases, the periodically oscillated environmental 
variable is the external temperature and its variations have been proved to 
induce macroscopic volumetric expansion and contraction cycles 
which can induce irreversible deformations in granular systems 
such as sand, silts and clays~\cite{RottaLoriaGEE2021}.
While the microscopic origins of the macroscopic response remains 
somehow elusive~\cite{PanGM2024}, X‐ray microtomography has revealed 
already that interactions happening at the particle level are key:
the material's thermally induced deformations (e.g. compaction) 
are strongly dependent on particles shape~\cite{PanGM2024, PanSR2024}, 
as well as on relative density and the prescribed temperature amplitude 
itself~\cite{RottaLoriaGEE2021, PanSR2024}.
Conceptually, it is not difficult to accept that subsequent periods
of expansion and contraction of the granular material  would lead, at 
least at a mesoscopic length scale, to 
a modulation in the region-to-region elasticity propagator, which is 
what our modeling proposes in a simplified approach to the complex 
soil material.

There is a deep connection between the sub-critical flow mechanism described 
here and cyclic fatigue in material science~\cite{Suresh1998}.
In fact, the cyclic fatigue phenomenon refers typically to the systematic 
increase in the length of micro-cracks by a fixed 
amount\footnote{In service conditions, the progressive increase in crack 
length produces an increase in stress intensity factors at the crack tip 
that lead eventually to a catastrophic failure of the sample} 
at every cycle of increase and decrease of the stress applied to a sample. 
There is a strong analogy between this process and the finding of a constant 
increase $\Delta X$ under increase and decrease of spring constants in our case.
One important difference is that cracks do not form in our set up, since 
the detaching of particles from their potential wells is followed by the 
re-attaching to a new well at a different position.
A second difference with the fatigue scenario is that, there,
a (quasistatic) temporal variation of the applied stress is applied, 
and there are no changes in intrinsic parameters of the model.
Notice that the finding of a well defined lower endurance limit for the 
oscillatory deformation in the present case (namely, a non-zero value 
of $\sigma_0$ or $f_0$) is very similar to the finding of a 
{\em fatigue limit} for crack propagation\footnote{Also known as 
{\em intrinsic strength} for polymeric materials, 
see~\cite{RobertsonInBook2021, BhowmickJMS1988}.} in some materials,
below which cracks do not propagate at all~\cite{MarghituMEH2001}.
To further extend this analogy, in future work we plan to explore cases 
where the perturbation takes 
the form of an externally oscillating stress, either aligned with or 
in a different direction from the average stress.

Results on shear-oscillated granular 
systems~\cite{YuanNatComm2024} showing that
particle roughness on a given length-scale 
could effectively affect the energy landscape and 
facilitate flow below the expected critical amplitude
could constitute interesting analogous cases of critical 
threshold depletion. 
On the theoretical side, it would be interesting to analyze the
mechanically stable configurations both below and above $f_0$ 
during the oscillatory protocol in the context of the Edwards 
thermodynamics~\cite{GradenigoPRL2015}.

\section{Summary and conclusions}
\label{sec:discussion}

In this work, we investigated a mechanism for athermal, sub-critical material flow 
driven by periodic variations in a parameter that affects internal structural 
forces in an externally driven system.
We illustrated this mechanism using a minimal model: two particles connected by 
a spring of variable stiffness $k$, which undergoes reptation down an inclined 
plane with a finite displacement $\Delta X$ per cycle of periodic variation in $k$.
We then extended this oscillatory mechanism to spatially distributed models of 
depinning and yielding transitions.

We demonstrated that when the external driving force $f$ is below the critical 
threshold  $f_c$ required for a steady deformation with time at a finite rate
($dX/dt>0$), there is a regime in which the system exhibits synchronized evolution
with the periodic variation of $k$, which represents the global elastic rigidity.
The deformation per cycle, $\Delta X$, decreases as $f$ is reduced and vanishes 
at a well-defined threshold, $f_0$.
This results was obtained numerically and also analytically in a mean field 
version of the problem.

The discovery of a sharp $f_0$ value that separates a long-lasting evolving 
regime from a non-evolving one is particularly remarkable.
This behavior is fundamentally different from thermal creep, where thermal 
activation always induce a finite creep rate, even at arbitrarily low $f$
(although vanishingly small as $f$ is reduced).
Furthermore, the similarity between the behavior of $\Delta X$ near $f_0$
under variation of $k$ and that of $v$ near $f_c$ for fixed $k$ suggests 
that the system may exhibit criticality at $f_0$, analogous to its critical 
behavior at $f_c$.
Our analysis of the elastic manifold’s roughness near $f_0$ revealed a 
divergence in the interface width $w$, a key indicator of criticality.
Current results suggests that the critical exponents at $f_0$
may be the same as those at $f_c$, though further detailed numerical 
analysis is required to confirm this.

Our findings have potential implications for interpreting geophysical 
processes at the Earth's surface.
While the persistent downhill creep of natural soils remains a subject 
of debate, laboratory experiments suggest that environmental disturbances 
play a crucial role.
In particular, Deshpande and co-workers~\cite{DeshpandeNC2021} assign to 
daily temperature fluctuations the ability to `rejuvenate' the sandpiles 
through thermomechanical stresses and sustain an approximately constant 
creep rate through repeated heating and cooling cycles.
This phenomenon is not yet fully understood, but we believe to be widespread, 
extending beyond specific materials and experimental setups.

The mechanism we propose involves periodic variations in internal 
parameters that modulate inter-particle or inter-regional forces, 
which facilitates the system reptation or flow.
The timescales associated with this sub-critical displacement are 
tightly coupled to the period of parameter oscillation, suggesting 
that similar mechanisms could operate in natural environments, 
linked to daily, seasonal, or even geological-scale cyclic variations.
Our modeling approach, where an interaction spring constant changes 
in an adiabatic manner, provides a simplified but well-founded 
framework for capturing sub-critical flow driven by environmental 
conditions such as temperature and humidity.
Our hypothesis, supported indirectly by observations in thermally cycled 
granular systems, posits that periodic environmental changes induce periodic 
oscillations in the system’s effective internal parameters.
We demonstrated that this effect is relevant in models of driven elastic 
interfaces in disordered media and in models of amorphous solids under deformation, 
revealing a regime of externally driven sub-critical flow that remains entirely 
athermal.
The connection between environmental variability and internal elasticity 
is key and warrants further systematic study.
Furthermore, our generic results suggest looking for particular models to 
describe specific types of oscillatory perturbations, such as:
particles with quasi-static size oscillations (e.g., due to daily or seasonal 
thermal expansion) as in~\cite{JaglaSM2023},
clays with adhesion properties modulated by humidity changes,
and systems with oscillating confinement geometries 
(e.g., periodically moving lateral boundaries).
While the specifics may vary, we expect the underlying physical mechanism 
to remain qualitatively the same.

\appendix
\renewcommand\thefigure{\thesection.\arabic{figure}}    

\section{Simulation details}
\label{app:methods}

\setcounter{figure}{0}    

\begin{figure}[b!]
\includegraphics[width=0.9\columnwidth]{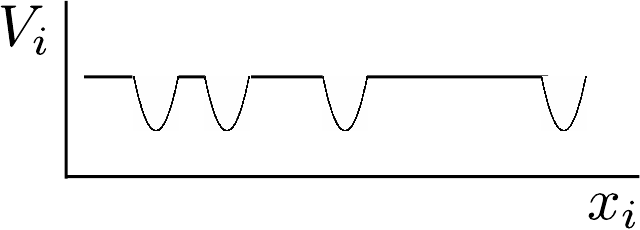}
\caption{
Schematic representation of the typical disorder potential we use. 
All parabolic wells are identical.
Inter-wells segment lengths are taken randomly from an exponential distribution.}
\label{fig:disordered_potential}
\end{figure}

In this Appendix we provide some more details regarding the simulations and 
parameters used.
The models used for both depinning and yielding transitions have been presented 
in Sec.~\ref{sec:framework}.

One specification to be made is that when working with long-range interactions,
i.e., in the elastoplactic model for the yielding transition, we make use of a 
pseudospectral method.
This is, the Eshelby kernel is defined in Fourier space as

\begin{equation}
G^{Y}_{\bf q}=k\frac{(q_x^2-q_y^2)^2}{(q_x^2+q_y^2)^2}
\end{equation}
and from here the precise form in real space is obtained (see Eq.~\ref{eq:eshelby_real}).
Then, at each step of the dynamics, the strain field appearing in Eq.~\ref{eq:eom_yielding}
is converted to Fourier space and convoluted with the kernel. The result is 
anti-transformed to get back the elastic interactions in real space.

For the disorder potential energy $V(x)$ appearing both in Eqs.~\ref{eq:eom_depinning} 
and \ref{eq:eom_yielding} we have adopted a function which alternates between parabolic 
wells and flat regions, as schematically depicted in Fig.~\ref{fig:disordered_potential}.
All parabolas are taken to be identical, defined by a unitary 
curvature and unitary width between the starting and ending points of the wells. 
The inter-wells flat regions, instead, are of different lengths, taken randomly
from an exponential distribution, uncorrelated from site to site.
This is the element that introduces randomness in the model.
We have also tested other types of disorder potentials, as for example 
the direct concatenation of parabolic wells of different sizes used in previous 
works~\cite{FerreroSM2019, FerreroPRM2021}.
The observed physics does not change qualitatively, but depending on the parameters
the sub-critical reptation region can be very narrow and visible only very close 
to $f_c$. 
The intercalation of wells and flat regions somehow helps the elastic manifold 
systems to enhance the oscillatory creep effect.
As a matter of fact, notice that the limit in which the parabolic wells become 
very narrow `traps', the local strain increase is purely plastic 
corresponds to the case in which we can build 
the equivalence between the elastic manifolds depinning-like models and the 
classic elastoplastic models of amorphous solids; these typically use a binary `state'
(elastic/plastic) variable for the model building blocks along with the local
stress~\cite{NicolasRMP2018}.
Therefore our $V(x)$ choice lays in between the ones typically used for 
depinning and for yielding and serves well to show the sub-critical reptation 
effect in both cases.

For a given fix value of $k=k_L$ in Eqs.~\ref{eq:eom_depinning} and
\ref{eq:eom_yielding}-\ref{eq:eshelby_real} and $f>f_c$ one can reach a 
stationary state after a transient by running a simulation for a moderate time, 
depending on the initial configuration.
If the starting condition corresponds to a steady state configuration obtained 
for a slightly larger force, the new steady state is reached very 
fast, typically a few hundreds time steps.
In fact, that is what we do to obtain the flowcurves of 
Figs.~\ref{fig:flowcurves_depinning} and \ref{fig:flowcurves_yielding}: 
we start at a large force, reach a steady state there and then 
slightly decrease the force and run stabilization periods at each 
step to take measurements in the steady states.

\begin{figure}[t!]
\includegraphics[width=\columnwidth]{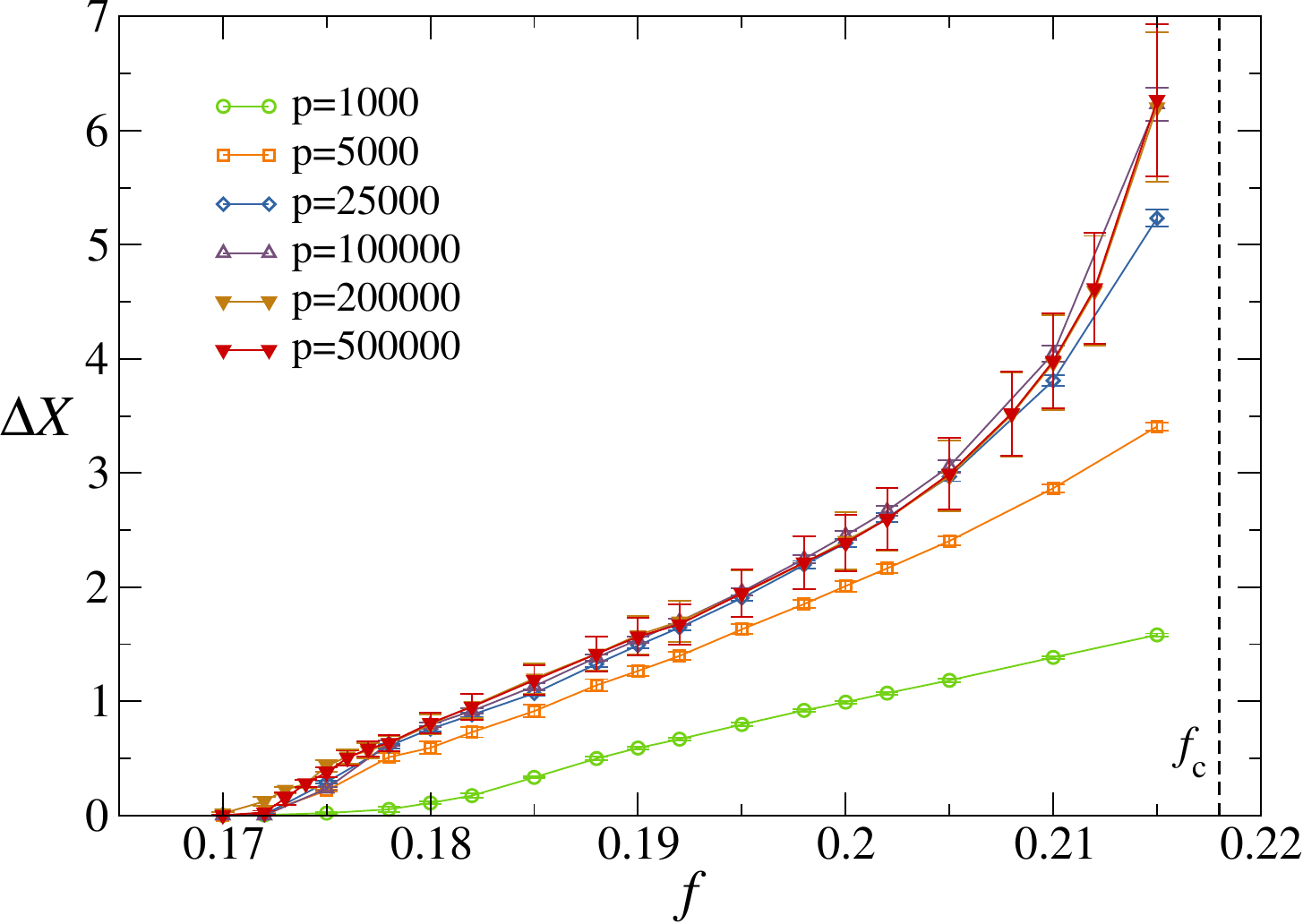}
\caption{
\textit{2D short-ranged depinning model.}
Dependency of the estimated value of $\Delta x$ on the period number $p$. 
System size is $N=512\times 512$, $k_L=0.1$, $k_S=0$.
}\label{fig:DX_varying_period_dep}
\end{figure}

On the contrary, reaching a steady state in the oscillatory protocol 
at $f<f_c$ is not that computationally cheap.
First, we have noticed that there is a strong dependency of the obtained
values of $\Delta X$ (eventually, even of $f_0$) with the frequency 
(equiv. period) of oscillation.
Fig.~\ref{fig:DX_varying_period_dep} shows the frequency dependence
of the average advance per oscillation period $\Delta x$ in the case 
of the depinning model.
If we want to work in an adiabatic limit, the pass of change of $k$
should be small enough to have results that are independent on it.
Trying to catch such a quasistatic limit, we have chosen the transition 
between $k_S$ and $k_L$ to be very slow. 
We have found both for depinning and yielding that a period of $20000$ 
steps was enough to guarantee frequency-independence in our results within 
error estimations of $\Delta x$ for most forces, and therefore used that 
value along the study. 
Nevertheless, for different system sizes, and in particular very close 
to $f_c$, this quantity should be adjusted to reach a frequency-independent 
steady state value.
Secondly, once the period is defined, one needs to run a large number of 
those for $\Delta x$ and $W$ to actually stabilize a mean value.
Typically we use a transient of 500 periods that we discard 
to reach the steady state and then yet another 500 periods to 
take measurements and averages.
So, large slow periods and many of them are needed to build
Figs.~\ref{fig:Dg_sigma_yielding}, \ref{fig:DX_vs_f_depinning_varying_ks}, 
\ref{fig:w_depinning} and \ref{fig:w_yielding}. 
That's why we have restricted ourselves to small and moderate system sizes.

\section{Analytical results in the mean field model}
\label{app:analyticMF}

\setcounter{figure}{0}    

Many details of the mean field model can be worked out analytically.
We describe here the kind of treatment that is necessary for these 
calculations, and present a few results.
In particular, we show the existence of a range $f_0$-$f_c$ in which there 
is a cyclic advance of the system upon oscillation of the value of $k$, 
and workout the value of $\Delta X$ in this range of applied forces.

As described before, the system consists of $N$ particles that move in 
a one-dimensional axis $x$ under the action of a potential consisting of a 
collection of very narrow wells, randomly distributed along $x$ 
(with a mean separation $a$, this implies an exponential distribution of 
inter-well distances).
Potential wells have a maximum pinning force that they are able to 
withstand, that we call $f_p$, and is the same for all the wells. 
A particle in a stationary situation can be located within a well 
(as long as the force acting on it is lower than $f_p$) or in the region between 
two wells. 
In this last case, the position of the particle is determined by the condition 
that the total force acting on it must be zero.

\begin{figure}[t!]
\includegraphics[width=1.0\columnwidth]{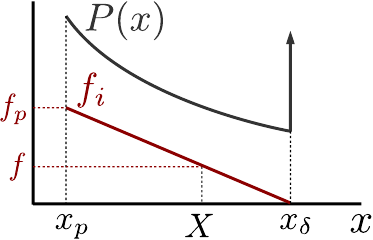}
\caption{
Schematic representation of the probability distribution function $P(x)$ 
for a system at the critical force $f_c$.
Note that $P(x)$ consists of the continuous black line plus the delta 
peak at the right.
The red line shows the $x$-dependent force on the particles $f_i$, which
is equal to $f_p$ at the left-most point of the distribution, zero at 
the delta peak, and the externally applied force $f$ at $X$. 
}\label{pdex}
\end{figure}

In the present mean field representation the force acting on particle $i$ is
\begin{equation}
f_i=k(X-x_i)+f
\label{eq:fi_mf}
\end{equation}
with $X=N^{-1}\sum_i x_i$ being the average coordinate position of 
the system. Note that this force is linear in $x_i$ (see Fig.~\ref{pdex}). 
Given any initial condition, upon setting a global $f>0$, all particles advance 
to the right until an equilibrium is reached. 
This occurs when every particle has either reached a well from which it cannot 
scape and $f_i<f_p$, or on its path to the next well has reached the position 
with $f_i=0$, and stays there. 
The position $x_{\delta}$ where this happens has to adjust to the 
condition $f+k(X-x_\delta)=0$.
Therefore, particles that are outside wells are all located at this 
same position.
Introducing the function $P(x)$ that gives the probability 
distribution of finding a particle at coordinate position $x$, 
a bit of analysis leads to the conclusion that the 
$P(x)$ consist of an exponential piece 
(originated in the exponential distribution of the inter-well distances), 
plus the delta peak populated by the particles that are outside 
wells (see Fig.~\ref{pdex}).
The exponential part of $P(x)$ starts at the point defined as $x_p$, 
where the force equals $f_p$, and extends to the point $x_\delta=X+f/k$ 
where the force is zero and where the delta peak occurs.
In other words, at fixed $k$ and $f$, particles outside wells are always 
the most advanced ones. 

\begin{figure}[t!]
\includegraphics[width=0.9\columnwidth]{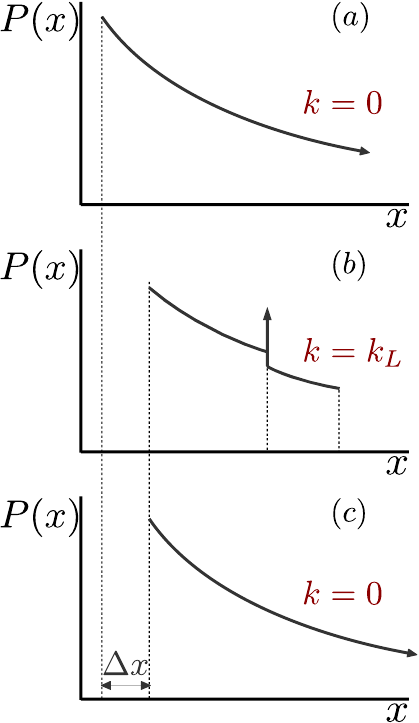}
\caption{
Schematic evolution of $P(x)$ for an applied force $f$ ($f_0<f<f_c$). 
(a) Distribution at $k=k_S\equiv 0$. This is a purely exponential distribution. 
(b) Distribution at $k=k_L$. Some of the right-most particles in (a) have jumped back 
to the position where $f_i=0$. Others from the left-most part have move to the right. 
(c) Distribution when $k$ is set back to $k_S\equiv 0$. The distribution is similar to 
that in (a), but displaced to the right a distance $\Delta X$.}
\label{pdex2}
\end{figure}

Let's consider the situation of a system driven just above 
the critical force: $f=f_c^+$. 
When $f \lesssim f_c$, the most retarded particles (namely most to the left 
along the $x$ axis) are trapped in a well and supporting the largest elastic force. 
In the continuous evolution of the system, when $f$ just overcomes $f_c$ and a 
finite global velocity is set, those particles are the first to jump out of 
their wells, given that all wells has the same pinning force $f_p$.
When they jump out, they reach the next well to the right, or stay at 
the point where $f_i=0$ if this happens before (to the left of) the 
next well.
Note that in the present case Eq.~\ref{eq:fi_mf} implies $f_p-f_c=(X-x_p)k$.
From its very definition, we also have $X=\int xP(x)dx$. 
Introducing the form of $P(x)$ and after a bit of manipulation and combination 
with the previous expression we obtain the value of the critical force as

\begin{equation}
f_c=f_p(1-z+z e^{1/z})
\end{equation}
with $z\equiv ka/f_p$.

Let us now analyze the case in which a value $f<f_c$ is applied 
and the value of $k$ is cycled quasi-statically between a large ($k_L$) 
and a small value ($k_S$). 
For simplicity we describe the situation when $k_S$ is zero (Fig.~\ref{pdex2}).
Provided an initial condition in equilibrium at a non-zero applied 
force $f<f_c$ pointing to the right and $k=k_L$, 
we start from a $P(x)$ distribution similar to that of Fig.~\ref{pdex}.
Reducing the value of $k$ to zero does not affect the position 
of particles that are located inside wells, but those that were at the 
delta peak of $P(x)$ will see their free from the elastic force that 
was holding them an drift towards the right until they find a new
potential well. 
In the end, for $k=0$ the $P(x)$ distribution becomes a pure exponential 
$P(x)\sim e^{-x}$ starting at some given point $x_0$. 
This is indicated schematically in Fig.~\ref{pdex2}(a).
Note also that from the history of the dynamical evolution, given a particle 
located at a well at $x_i$, we can be sure there is no other well for that 
particle in the interval $(x_0,x_i)$.
Now, when the value of $k$ is increased again ($k\to k_L$), the elastic 
force start to act on the particles.
All those on the right of the mean value $X$ will feel a force pushing then 
to the left, and for those with the largest values of $x$ such force would
overcome $-f_p$ and they will jump out of their wells but now towards the left.
Because of the previous comment, those particles do not reach a new well but
regain a position in which $f_i=0$ for such $k=k_L$, creating a delta
peak at a position that now is intermediate in $P(x)$
(those particles feeling a force $0>f>-f_p$ persist in their wells
and have $x>x_\delta$).
Along this process, it may happen (and it happens eventually, i.e., when $f>f_0$) 
that some of the left-most particles receive a positive force larger that $f_p$, 
and they jump to the right (dragged by a mean value $X$ that has moved forward 
in the previous step). 
The final distribution at $k=k_L$ is qualitatively seen in Fig.~\ref{pdex2}(b). 
Finally, when $k$ is turned to zero again, the process is repeated but 
with some particles already advanced respect to the previous cycle.
We obtain the result in Fig.~\ref{pdex2}(c), namely a distribution similar 
to the one in Fig.~\ref{pdex2}(a), but displaced to the right in an amount $\Delta X$.

Based in the  qualitative evolution just mentioned, it is possible to calculate the 
value of $\Delta X$ given the value of $k_L$. 
The calculation is elemental, but a bit cumbersome.
The outcome is the following.
First, one calculates $\tilde z$ from
\begin{equation}
f/f_p=1-\tilde z+(1+\tilde z)e^{-2/\tilde z}.
\end{equation}
Then, $\Delta X$ is calculated from
\begin{equation}
f/f_p=1- z+z e^{-1/z}(1-e^{-\Delta X}) + (2z/\tilde z -1+z-z\Delta X) e^{-2/\tilde z}.
\end{equation}
The obtained $\Delta X(f)$ curve is plotted in the left panel of 
Fig.~\ref{fig:delta_x_medio}.
In particular, setting $\Delta X$ to zero provides the minimum value $f_0$ 
necessary to observe the advance of the system upon oscillation of $k$, which is
\begin{equation}
f_0=f_p(1-z+(1+z)e^{-2/z}).
\end{equation}
This dependence was shown in Fig.~\ref{fig:s1s2}.

\begin{acknowledgments} 
We acknowledge support from PIP 2021-2023 CONICET Project Nº~757
and the CNRS IRP Project ``Statistical Physics of Materials''.
EEF acknowledges support from the Maria Zambrano program of the 
Spanish Ministry of Universities through the University of Barcelona,
and MCIN/AEI support through PID2019-106290GB-C22.
\end{acknowledgments}

%

\end{document}